\documentclass[twocolumn]{autart}    
\usepackage[cmex10]{amsmath}
\numberwithin{equation}{section}
\usepackage{amssymb}
\usepackage{mathdots}
\usepackage{psfrag}
\usepackage{verbatim}
\usepackage{nameref}

\usepackage{lineno,hyperref}
\modulolinenumbers[5]
\usepackage{natbib}

\usepackage{mathrsfs}
\usepackage{mathtools}
\usepackage{mathdots}
\usepackage{graphicx}          
\usepackage{enumitem}
\newcounter{muni}
\newenvironment{remunerate}
               {\begin{list}{{\upshape 
               \arabic{muni}.}}{\usecounter{muni}
                \setlength{\leftmargin}{0pt}
                \setlength{\itemindent}{25pt}}}{\end{list}}

\makeatletter
\newcommand{\labitem}[2]{%
\def\@itemlabel{#1}
\item
\def\@currentlabel{#1}\label{#2}}
\makeatother
\newcommand{\overbar}[1]{\mkern 1.5mu\overline{\mkern-1.5mu#1\mkern-1.5mu}\mkern 1.5mu}

\newenvironment{smallarray}[1]
 {\null\,\vcenter\bgroup\scriptsize
  \arraycolsep=.2em
  \hbox\bgroup$\array{@{}#1@{}}}
 {\endarray$\egroup\egroup\,\null}

\def\smallint{\begingroup\textstyle \int\endgroup}

\begin{document}
\setlength{\parskip}{6pt}

\begin{frontmatter}

\title{On reciprocal systems and controllability}

\author[thh22]{Timothy H.\ Hughes}\ead{t.h.hughes@exeter.ac.uk}    
\address[thh22]{Department of Mathematics, University of Exeter, Penryn Campus, Penryn, Cornwall, TR10 9EZ, UK}
\thanks{\copyright \hspace{0.1cm} 2018. This manuscript is made available under the CC-BY-NC-ND 4.0 license \url{http://creativecommons.org/licenses/by-nc-nd/4.0/}}
          

\begin{abstract}                          
In this paper, we extend classical results on (i) signature symmetric realizations, and (ii) signature symmetric and passive realizations, to systems which need not be controllable. These results are motivated in part by the existence of important electrical networks, such as the famous Bott-Duffin networks, which possess signature symmetric and passive realizations that are uncontrollable. In this regard, we provide necessary and sufficient algebraic conditions for a behavior to be realized as the driving-point behavior of an electrical network comprising resistors, inductors, capacitors and transformers.
\end{abstract}

\end{frontmatter}

\newtheorem{example}[thm]{Example}
\providecommand{\abs}[1]{\lvert#1\rvert}
\newtheorem{thmapp}{Theorem}[section]
\newtheorem{lemapp}[thmapp]{Lemma}
\newtheorem{defapp}[thmapp]{Definition}
\newtheorem{remapp}[thmapp]{Remark}

\section{Introduction}
This paper is concerned with reciprocal systems \citep[see, e.g.,][]{CAS_RT, JWDSP2, AndVong, newclms, VDSGS}. Reciprocity is an important form of symmetry in physical systems which arises in acoustics (Rayleigh-Carson reciprocity); elasticity (the Maxwell-Betti reciprocal work theorem); electrostatics (Green's reciprocity); and electromagnetics (Lorentz reciprocity), where it follows as a result of Maxwell's laws \cite[p.\ 43]{newclms}. Special cases of reciprocal systems include reversible systems, as arise in thermodynamics; and relaxation systems, such as viscoelastic materials \citep{JWDSP2}. In addition, reciprocity is a property of important classes of electrical, mechanical and structural systems, such as lightly damped flexible structures \citep{NIS1}. Our focus in this paper is on linear reciprocal systems. In contemporary systems and control theory, a linear reciprocal system is typically defined as a system with a symmetric transfer function. A fundamental result in systems and control theory states that if the transfer function is also proper, then the system possesses a so-called \emph{signature symmetric realization} \citep[see][]{JWDSP2, AndVong, PFSRTF, YoulaTissi}. However, this result is subject to one notable caveat: the system is assumed to be controllable.

Practical motivation for developing a theory of reciprocity that does not assume controllability arises from electrical networks. Notably, the driving-point behavior of an electrical network comprising resistors, inductors, capacitors and transformers (an RLCT network) is necessarily reciprocal, and also passive,\footnote{A system is passive if the net energy that can be extracted from the system into the future is bounded above (this bound depending only on the past trajectory of the system).} but it need not be controllable \citep[see][]{camwb, JWHVDS, HugGMF}. Indeed, as noted by \citet{camwb}, it is not known what (uncontrollable) behaviors can be realized as the driving-point behavior of an RLCT network. In addition, an RLCT network need not possess an impedance function, so the conventional definition of a reciprocal system as one with a symmetric transfer function is inappropriate for such networks.

The purpose of this paper is to address the aforementioned limitations with the theory of reciprocity. The paper is structured as follows. In Section \ref{sec:ctr}, we review the classical theory of reciprocal systems in more detail, with a particular focus on passive and reciprocal systems, to highlight the limitations of the existing theory and the contributions of this paper. In Sections \ref{sec:recip} and \ref{sec:rssr}, we proceed to develop a theory of reciprocal systems which addresses these limitations. Sections \ref{sec:prb} and \ref{sec:raps} are then concerned with systems that are both reciprocal and passive, such as RLCT networks. The main results are summarised in the following two paragraphs. 

In Section \ref{sec:recip}, we provide a formal definition of reciprocity (Definition \ref{def:rb}), which was first proposed by \cite{newclms}. The main advantage of this definition is that it does not assume the existence of a symmetric transfer function. This is particularly fitting in the context of electrical networks as these need not possess an impedance function. We then provide a 2-part theorem which we call the \emph{reciprocal behavior theorem}. In part 1 (Theorem \ref{thm:sigsymrep}), we provide necessary and sufficient conditions for a system to be reciprocal in terms of the differential equations describing the system. We also prove that, for any given reciprocal system, it is possible to permute the system's variables to obtain a system with a proper symmetric transfer function. Part 2 (Theorem \ref{thm:ssssr}) then proves the existence of a signature symmetric realization for any given system with a proper symmetric transfer function (irrespective of controllability). 

Section \ref{sec:prb} contains another 2-part theorem: the \emph{passive and reciprocal behavior theorem}. Part 1 (Theorem \ref{thm:prbtp1}) provides necessary and sufficient algebraic conditions for a system to be passive \emph{and} reciprocal in terms of the differential equations describing the system. This theorem also answers the first open problem posed in \cite{camwb} in the more general setting of multi-port networks: it is shown that a behavior $\mathcal{B}$ is realizable as the driving-point behavior of an RLCT network if and only if $\mathcal{B}$ is passive and reciprocal. Part 2 (Theorem \ref{thm:prprop}) then proves the existence of a passive and signature symmetric realization for any given passive system with a proper symmetric transfer function. The results in this section build on earlier results in \citep{THTPLSNA, THAS} on systems which are passive but not necessarily reciprocal. The extension to consider passive \emph{and} reciprocal systems is by no means trivial, and depends on a number of supplementary lemmas that are provided in Section \ref{sec:raps} and Appendix \ref{sec:prbtsl}. Finally, the proofs in the paper, together with existing results in the literature, provide an algorithm for constructing an RLCT network realization of an arbitrary given reciprocal and passive behavior. This is illustrated by two examples in Section \ref{sec:ex}.

\section{Notation and Preliminaries}
\label{sec:not}
We denote the real and complex numbers by $\mathbb{R}$ and $\mathbb{C}$, and the open and closed right-half plane by $\mathbb{C}_{+}$ and $\overbar{\mathbb{C}}_{+}$. If $\lambda \in \mathbb{C}$, then $\bar{\lambda}$ denotes its complex conjugate. The polynomials, rational functions, and proper (i.e., bounded at infinity) rational functions in the indeterminate $\xi$ with real coefficients are denoted $\mathbb{R}[\xi], \mathbb{R}(\xi)$, and $\mathbb{R}_{p}(\xi)$. The $m {\times} n$ matrices with entries from $\mathbb{R}$ (resp., $\mathbb{R}[\xi]$, $\mathbb{R}(\xi)$, $\mathbb{R}_{p}(\xi)$) are denoted $\mathbb{R}^{m \times n}$ (resp., $\mathbb{R}^{m \times n}[\xi]$, $\mathbb{R}^{m \times n}(\xi)$, $\mathbb{R}_{p}^{m \times n}(\xi)$), and $n$ is omitted if $n=1$. We denote the block column and block diagonal matrices with entries $H_{1}, \ldots , H_{n}$ by $\text{col}(H_{1} \hspace{0.15cm} \cdots \hspace{0.15cm} H_{n})$ and $\text{diag}(H_{1} \hspace{0.15cm} \cdots \hspace{0.15cm} H_{n})$; and we will use horizontal and vertical lines to indicate the partition in block matrix equations (e.g., see (\ref{eq:cpf1})). If $H \in \mathbb{R}^{m \times n}, \mathbb{R}^{m \times n}[\xi]$, or $\mathbb{R}^{m \times n}(\xi)$, then $H^{T}$ denotes its transpose, and if $H$ is nonsingular (i.e., $\det{(H)} \neq 0$) then $H^{-1}$ denotes its inverse. If $H \in \mathbb{R}^{m \times n}$, then $\text{rank}(H)$ denotes its rank; and if $G \in \mathbb{R}^{m \times n}(\xi)$, then $\text{normalrank}(G) \coloneqq \max_{\lambda \in \mathbb{C}}(\text{rank}(G(\lambda)))$. If $M \in \mathbb{R}^{m \times m}$, then $\text{spec}(M) \coloneqq \lbrace \lambda \in \mathbb{C} \mid \text{det}(\lambda I {-} M) = 0\rbrace$; and if, in addition, $M$ is symmetric, then $M > 0$ ($M \geq 0$) indicates that $M$ is positive (non-negative) definite. A matrix $\Sigma \in \mathbb{R}^{n \times n}$ is called a signature matrix if it is diagonal and all of its entries are either $1$ or $-1$. A $V \in \mathbb{R}^{n \times n}[\xi]$ is called unimodular if $\det{(V)}$ is a non-zero constant (equivalently, $V$ is nonsingular with $V^{-1} \in \mathbb{R}^{n \times n}[\xi]$). If $H \in \mathbb{R}^{n \times n}(\xi)$, then $H$ is called positive-real if $H$ is analytic in $\mathbb{C}_{+}$ and $H(\bar{\lambda})^{T} + H(\lambda) \geq 0$ for all $\lambda \in \mathbb{C}_{+}$. 

The ($k$-vector-valued) locally integrable functions are denoted $\mathcal{L}_{1}^{\text{loc}}\left(\mathbb{R}, \mathbb{R}^{k}\right)$ \citep[Defns.\ 2.3.3, 2.3.4]{JWIMTSC}, and we equate any two locally integrable functions that differ only on a set of measure zero. The ($k$-vector-valued) infinitely differentiable functions with bounded support on the left (resp., bounded support on the right, bounded support) are denoted $\mathcal{D}_{+}\left(\mathbb{R}, \mathbb{R}^{k}\right)$ (resp.,  $\mathcal{D}_{-}\left(\mathbb{R}, \mathbb{R}^{k}\right)$,  $\mathcal{D}\left(\mathbb{R}, \mathbb{R}^{k}\right)$). The convolution operator is denoted by $\star$; i.e., if $\mathbf{w}_{1}, \mathbf{w}_{2} \in \mathcal{D}_{+}\left(\mathbb{R}, \mathbb{R}^{k}\right)$, then $(\mathbf{w}_{1} \star \mathbf{w}_{2})(t) = \smallint_{-\infty}^{\infty}{\mathbf{w}_{1}(\tau)^{T}\mathbf{w}_{2}(t-\tau)d\tau}$.

A main contribution of this paper is to develop a theory of reciprocal systems which doesn't assume controllability, observability, or the existence of a transfer function. This is relevant to electric networks which can possess uncontrollable and unobservable internal modes, and whose driving-point currents and voltages need not adhere to the conventional system theoretic input-output view. The natural framework to formalise these issues is the behavioral approach \citep{JWIMTSC}. Accordingly, the remainder of this section contains relevant definitions and results on behaviors.

We consider behaviors (systems) defined as the set of weak solutions \citep[see][Section 2.3.2]{JWIMTSC} to a differential equation:
\begin{equation}
\hspace*{-0.3cm} \mathcal{B} = \lbrace \mathbf{w} \in \mathcal{L}_{1}^{\text{loc}}\left(\mathbb{R}, \mathbb{R}^{q}\right) \mid R(\tfrac{d}{dt})\mathbf{w} {=} 0\rbrace, \hspace{0.1cm} R \in \mathbb{R}^{p \times q}[s]. \label{eq:bd}
\end{equation}
The behavior $\mathcal{B}$ is called \emph{controllable} if, for any two trajectories $\mathbf{w}_{1}, \mathbf{w}_{2} \in \mathcal{B}$ and $t_{0} \in \mathbb{R}$, there exists $\mathbf{w} \in \mathcal{B}$ and $t_{1} \geq t_{0}$ such that $\mathbf{w}(t) = \mathbf{w}_{1}(t)$ for all $t \leq t_{0}$ and $\mathbf{w}(t) = \mathbf{w}_{2}(t)$ for all $t \geq t_{1}$ \cite[Definition 5.2.2]{JWIMTSC}. From \citep[Theorem 5.2.10]{JWIMTSC}, $\mathcal{B}$ in (\ref{eq:bd}) is controllable if and only if $\text{rank}(R(\lambda))$ is the same for all $\lambda \in \mathbb{C}$.

We pay particular attention to state-space systems:
\begin{align}
&\hspace*{-0.3cm} \mathcal{B}_{s} = \lbrace (\mathbf{u}, \mathbf{y}, \mathbf{x}) {\in} \mathcal{L}_{1}^{\text{loc}}\left(\mathbb{R}, \mathbb{R}^{n}\right) {\times} \mathcal{L}_{1}^{\text{loc}}\left(\mathbb{R}, \mathbb{R}^{n}\right) {\times} \mathcal{L}_{1}^{\text{loc}}\left(\mathbb{R}, \mathbb{R}^{d}\right) \mid \nonumber\\
& \hspace{1.0cm} \tfrac{d\mathbf{x}}{dt} = A\mathbf{x} + B\mathbf{u} \text{ and } \mathbf{y} = C\mathbf{x} + D\mathbf{u} \rbrace, \nonumber \\
&\hspace*{-0.3cm} A \in \mathbb{R}^{d \times d}, B \in \mathbb{R}^{d \times n}, C \in \mathbb{R}^{n \times d} \text{ and } D \in \mathbb{R}^{n \times n}. \label{eq:bhssr}
\end{align}
Here, we call the pair $(A,B)$ \emph{controllable} if $\mathcal{B}_{s}$ is controllable; and we call the pair $(C,A)$ \emph{observable} if $(\mathbf{u}, \mathbf{y}, \mathbf{x}) \in \mathcal{B}_{s}$ and $(\mathbf{u}, \mathbf{y}, \hat{\mathbf{x}}) \in \mathcal{B}_{s}$ imply $\mathbf{x} = \hat{\mathbf{x}}$ \cite[Definition 5.3.2]{JWIMTSC}. 
These concepts are equivalent to the well known algebraic conditions for controllability/observability of a pair of matrices \citep[see][Chapter 5]{JWIMTSC}.

We also consider behaviors obtained by transforming and/or eliminating variables in a behavior $\mathcal{B}$ as in (\ref{eq:bd}). For example, associated with the state-space system $\mathcal{B}_{s}$ in (\ref{eq:bhssr}) is the corresponding external behavior $\mathcal{B}_{s}^{(\mathbf{u}, \mathbf{y})} = \lbrace (\mathbf{u},\mathbf{y}) \mid \exists \mathbf{x} \text{ with } (\mathbf{u},\mathbf{y},\mathbf{x}) \in \mathcal{B}_{s}\rbrace$. More generally, if $T_{1} \in \mathbb{R}^{p_{1} \times q}, \ldots , T_{n} \in \mathbb{R}^{p_{n} \times q}$ are such that $\text{col}(T_{1} \hspace{0.15cm} \cdots \hspace{0.15cm} T_{n}) \in \mathbb{R}^{q \times q}$ is a nonsingular real matrix, and $m$ is an integer satisfying $1 \leq m \leq n$, then we denote the projection of $\mathcal{B}$ onto $T_{1}\mathbf{w}, \ldots , T_{m}\mathbf{w}$ by
\begin{multline*}
\hspace*{-0.4cm} \mathcal{B}^{(T_{1}\mathbf{w}, \ldots , T_{m}\mathbf{w})} = \lbrace (T_{1}\mathbf{w}, \ldots , T_{m}\mathbf{w}) \mid \exists (T_{m+1}\mathbf{w}, \ldots , T_{n}\mathbf{w}) \\ \text{such that } \mathbf{w} \in \mathcal{B}\rbrace.
\end{multline*}
A representation for the behavior $\mathcal{B}^{(T_{1}\mathbf{w}, \ldots , T_{m}\mathbf{w})}$ can be obtained by the so-called elimination theorem (see Appendix \ref{app:b}). In particular, by eliminating the state variables $\mathbf{x}$ from $\mathcal{B}_{s}$, we 
%
obtain a behavior of the form
\begin{align}
&\hspace*{-0.3cm} \hat{\mathcal{B}} {=} \lbrace (\mathbf{u}, \mathbf{y}) {\in} \mathcal{L}_{1}^{\text{loc}}\left(\mathbb{R}, \mathbb{R}^{n}\right) {\times} \mathcal{L}_{1}^{\text{loc}}\left(\mathbb{R}, \mathbb{R}^{n}\right) \mid \hat{P}(\tfrac{d}{dt})\mathbf{u} {=} \hat{Q}(\tfrac{d}{dt})\mathbf{y}\rbrace, \nonumber \\ 
&\hspace*{-0.3cm} \hat{P}, \hat{Q} \in \mathbb{R}^{n \times n}[\xi], \hat{Q} \text{ nonsingular and } \hat{Q}^{-1}\hat{P} \text{ proper}.\label{eq:bgsqd}
\end{align}
More specifically, from \citep[Sections 2 and 4]{THBRSF} we have the following lemma on behavioral realizations. 
\begin{lem}
\label{lem:ssr}
Let $\mathcal{B}_{s}$ be as in (\ref{eq:bhssr}) and $\mathcal{A}(\xi) {\coloneqq} \xi I {-} A$. There exist polynomial matrices $\hat{P},\hat{Q},Y,Z,U,V,E,F, G$ where
\begin{enumerate}[label=\arabic*., ref=\arabic*, leftmargin=0.5cm]
\item $\begin{bmatrix}Y& Z \\ U& V\end{bmatrix}\begin{bmatrix}-D& I& -C\\ -B& 0& \mathcal{A}\end{bmatrix} = \begin{bmatrix}-\hat{P}& \hat{Q}& 0\\ -E& -F& G\end{bmatrix}$; \label{nl:utsse1}
\item $\begin{bmatrix}Y& Z \\ U& V\end{bmatrix}$ is unimodular; and \label{nl:utsse2}
\item $G$ is nonsingular. \label{nl:utsse3}
\end{enumerate}

Furthermore, if conditions \ref{nl:utsse1}--\ref{nl:utsse3} hold and $\hat{\mathcal{B}}$ is as in (\ref{eq:bgsqd}), then $\mathcal{B}_{s}^{(\mathbf{u}, \mathbf{y})} = \hat{\mathcal{B}}$, and we say that $(A,B,C,D)$ is a realization of $(\hat{P},\hat{Q})$. Also, if $\hat{\mathcal{B}}$ is as in (\ref{eq:bgsqd}), then there exists $\mathcal{B}_{s}$ as in (\ref{eq:bhssr}) and polynomial matrices $Y,Z,U,V,E,F$ and $G$ satisfying conditions \ref{nl:utsse1}--\ref{nl:utsse3}.
\end{lem}

\begin{rem}
\label{rem:ssr}
\textnormal{For a given behavior $\hat{\mathcal{B}}$ as in (\ref{eq:bgsqd}), algorithms for computing a realization $(A,B,C,D)$ for $(\hat{P},\hat{Q})$ (i.e., a state-space system $\mathcal{B}_{s}$ such that $\mathcal{B}_{s}^{(\mathbf{u}, \mathbf{y})} = \hat{\mathcal{B}}$) are described in \citep[Section 4.7]{PFSOB} and \citep[Section 4]{THBRSF}. Such behavioral realizations are not unique. Indeed, it is easily shown from \citep[Note A.3]{THAS} that $(\hat{A},\hat{B},\hat{C},\hat{D})$ is another realization for $(\hat{P},\hat{Q})$ if and only if (i) $\hat{D} + \hat{C}(\xi I - \hat{A})^{-1}\hat{B} = D + C(\xi I - A)^{-1}B$; and (ii) there exist matrices $T_{1} \in \mathbb{R}^{\hat{d} \times d}$ and $T_{2} \in \mathbb{R}^{d \times \hat{d}}$ such that $CA^{i}T_{1} = \hat{C}\hat{A}^{i}$ for $i = 0, 1, 2, \ldots$, and $\hat{C}\hat{A}^{k}T_{2} = CA^{k}$ for $k = 0, 1, 2, \ldots$.\footnote{In fact, by the Cayley Hamilton theorem, it can be shown that these two conditions hold if and only if they hold for $i = 0, 1, \ldots , d$ and $k = 0, 1, \ldots , \hat{d}$.} Note that the equivalence of transfer functions (condition (i)) is necessary but not sufficient. E.g., let $B = 0$, $C = 1$ and $D = 1$, so $D + C(\xi I - A)^{-1}B = 1$ for all $A \in \mathbb{R}$. If $A = -1$, then $(u,y) \in \mathcal{B}_{s}^{(u,y)}$ if and only if there exists $k_{1} \in \mathbb{R}$ such that $y(t) = u(t) + k_{1}e^{-t}$. But if $A = 0$, then $(u,y) \in \mathcal{B}_{s}^{(u,y)}$ if and only if there exists $k_{2} \in \mathbb{R}$ such that $y(t) = u(t) + k_{2}$.}
\end{rem}

\section{Signature symmetric realizations of symmetric transfer functions}
\label{sec:ctr}
The following fundamental result in systems and control theory states that any given controllable system with a proper symmetric transfer function has a so-called \emph{signature symmetric realization}.
\begin{lem}
\label{thm:sigsr}
Let $\hat{\mathcal{B}}$ in (\ref{eq:bgsqd}) be controllable. Then the following are equivalent. 
\begin{enumerate}[label=\arabic*., ref=\arabic*, leftmargin=0.5cm]
\item $\hat{Q}^{-1}\hat{P}$ is symmetric.\label{nl:cssc1}
\item There exists $\mathcal{B}_{s}$ as in (\ref{eq:bhssr}) and a signature matrix $\Sigma_{i} \in \mathbb{R}^{d \times d}$ such that (i) $\hat{\mathcal{B}} = \mathcal{B}_{s}^{(\mathbf{u},\mathbf{y})}$; (ii) $(A,B)$ is controllable; (iii) $(C,A)$ is observable; and (iv) 
$A\Sigma_{i} = \Sigma_{i}A^{T}$, $\Sigma_{i}C^{T} = B$, and $D = D^{T}$.\label{nl:cssc2}
\end{enumerate}
\end{lem}

\begin{pf}
If $\hat{\mathcal{B}}$ in (\ref{eq:bgsqd}) is controllable, then there exists $\mathcal{B}_{s}$ as in (\ref{eq:bhssr}) which satisfies (i)--(iii) in condition \ref{nl:cssc2} \citep[see][Appendix D]{THTPLSNA}. Furthermore, $D + C(\xi I - A)^{-1}B = (\hat{Q}^{-1}\hat{P})(\xi)$, and it is then easily verified that if $A, B, C$ and $D$ are as in condition \ref{nl:cssc2} then $\hat{Q}^{-1}\hat{P}$ is symmetric. This proves that \ref{nl:cssc2} $\Rightarrow$ \ref{nl:cssc1}. The proof of \ref{nl:cssc1} $\Rightarrow$ \ref{nl:cssc2} then follows from \citep[Theorem 6]{JWDSP2} \citep[alternatively, see][]{AndVong, PFSRTF, YoulaTissi, RBA}. This proof proceeds by first showing that, if $\hat{A} \in \mathbb{R}^{d \times d}, \hat{B} \in \mathbb{R}^{d \times n}, \hat{C} \in \mathbb{R}^{n \times d}$ and $\hat{D} \in \mathbb{R}^{n \times n}$ are such that $\hat{D} + \hat{C}(\xi I - \hat{A})^{-1}\hat{B}$ is symmetric, $(\hat{A}, \hat{B})$ is controllable, and $(\hat{C}, \hat{A})$ is observable, then there exists a nonsingular symmetric $P \in \mathbb{R}^{d \times d}$ such that $P\hat{A} = \hat{A}^{T}P$, $\hat{C}^{T} = P\hat{B}$ and $\hat{D} = \hat{D}^{T}$. Note that, with the notation $\hat{V}_{o} = \text{col}(\hat{C} \hspace{0.15cm} \hat{C}\hat{A} \hspace{0.15cm} \ldots \hspace{0.15cm} \hat{C}\hat{A}^{d-1})$ and $\hat{V}_{c} = [\hat{B} \hspace{0.15cm} \hat{A}\hat{B} \hspace{0.15cm} \ldots \hspace{0.15cm} \hat{A}^{d-1}\hat{B}]$, then $P\hat{V}_{c} = \hat{V}_{o}^{T}$, whereupon $P$ can be computed from the explicit formula $P = \hat{V}_{o}^{T}\hat{V}_{c}^{T}(\hat{V}_{c}\hat{V}_{c}^{T})^{-1}$ \citep[Section 7.4]{AndVong}. Since $P$ is symmetric, then there exists a signature matrix $\Sigma_{i} \in \mathbb{R}^{d \times d}$ and a nonsingular $T \in \mathbb{R}^{d \times d}$ such that $P = T^{T}\Sigma_{i}T$. We then let $A := T\hat{A}T^{-1}, B := T\hat{B}, C := \hat{C}T^{-1}$ and $D := \hat{D}$.
\end{pf}

Of particular interest are controllable systems with proper symmetric transfer functions that are positive-real. These arise as the impedances of electrical networks containing resistors, inductors, capacitors and transformers (RLCT networks). In fact, such systems have a particular physical relevance, since every known physical system with a non-symmetric positive-real impedance actually contains active components \citep[see][]{NIS2}. A second fundamental result in systems and control theory is that any controllable system with a proper symmetric positive-real transfer function has a so-called \emph{passive and signature symmetric realization}, in accordance with the following lemma.
\begin{lem}
\label{thm:Grelssp}
Let $\hat{\mathcal{B}}$ in (\ref{eq:bgsqd}) be controllable. Then the following are equivalent:
\begin{enumerate}[label=\arabic*., ref=\arabic*, leftmargin=0.5cm]
\item $\hat{Q}^{-1}\hat{P}$ is positive-real and symmetric.
\item There exists $\mathcal{B}_{s}$ as in (\ref{eq:bhssr}) and a signature matrix $\Sigma_{i} \in \mathbb{R}^{d \times d}$ such that (i) $\hat{\mathcal{B}} = \mathcal{B}_{s}^{(\mathbf{u},\mathbf{y})}$; (ii) $(A,B)$ is controllable; (iii) $(C,A)$ is observable;
\begin{itemize}
\item[(iv)] $\begin{bmatrix}{-}A& {-}B\\ C& D\end{bmatrix}\! {+} \!\begin{bmatrix}{-}A& {-}B\\ C& D\end{bmatrix}^{T} \geq 0$; and \label{nl:crstc1}
\item[(v)] $A\Sigma_{i} = \Sigma_{i}A^{T}$, $\Sigma_{i}C^{T} = B$, and $D = D^{T}$.\label{nl:crstc2}
\end{itemize}
\end{enumerate}
\end{lem}

\begin{pf}
See \citep[Theorem 7]{JWDSP2}. 
\end{pf}

\begin{figure}[!b]
\scriptsize
 \begin{center}
\leavevmode
\includegraphics[width=0.8\hsize]{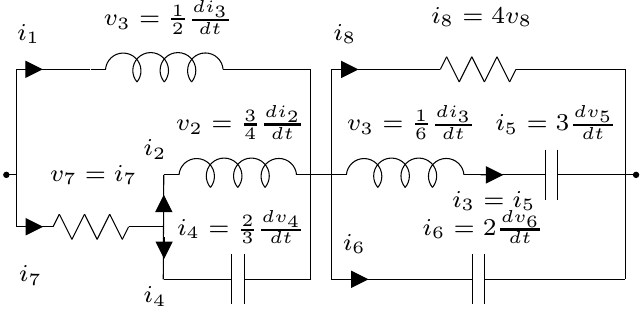}
\end{center}
\caption{Bott-Duffin realization of the driving-point behavior $(\tfrac{d}{dt}+1)(\tfrac{d^2}{dt^2}+\tfrac{d}{dt}+1)i = (\tfrac{d}{dt}+1)(\tfrac{d^2}{dt^2}+\tfrac{d}{dt}+4)v$}
\label{fig:bdne}
\end{figure}
Using the reactance extraction approach, any realization of the form of Lemma \ref{thm:Grelssp} gives rise to an RLCT network whose impedance is equal to $\hat{Q}^{-1}\hat{P}$ \citep[see][]{AndVong}. However, Lemma \ref{thm:Grelssp} contains several notable assumptions that are not satisfied by many RLCT networks. First, the theorem assumes the existence of a proper symmetric transfer function, yet not all RLCT networks possess a proper impedance \citep[see][Section 3]{THTPLSNA}. 
Second, the theorem assumes the system is controllable, but not all RLCT networks have controllable driving-point behaviors. Examples include the famous Bott-Duffin networks and their simplifications \citep[see][]{HugSmSP, HugNa, HugGMF}. One such network is provided in Fig.\ \ref{fig:bdne}, whose behavior is described by the state-space realization
\begin{align*}
\frac{d\mathbf{x}}{dt}
&= \left[\!\begin{smallmatrix}-2& 0& 0& -\sqrt{3}& 0& 0\\0 & 0& 0& -\sqrt{2}& 0& 0\\ 0& 0& 0& 0& \sqrt{2}& -\sqrt{3}\\ \sqrt{3}& \sqrt{2}& 0& 0& 0& 0\\ 0& 0& -\sqrt{2}& 0& 0& 0\\ 0& 0& \sqrt{3}& 0& 0& -2\end{smallmatrix}\!\right]\mathbf{x}
+ \left[\!\begin{smallmatrix}-\sqrt{2}\\ 0\\ 0\\ \sqrt{\tfrac{3}{2}}\\ 0\\ \tfrac{1}{\sqrt{2}}\end{smallmatrix}\!\right]i\\
v &= \left[\!\begin{smallmatrix}-\sqrt{2}& 0& 0& \sqrt{\tfrac{3}{2}}& 0& \tfrac{1}{\sqrt{2}}\end{smallmatrix}\!\right]\mathbf{x} + i, \text{ where} \\ 
\mathbf{x} &= \text{col}(\tfrac{v_{1}}{\sqrt{2}} \hspace{0.15cm} \tfrac{\sqrt{3}}{2}v_{2} \hspace{0.15cm} \tfrac{v_{3}}{\sqrt{6}} \hspace{0.15cm} \sqrt{\tfrac{2}{3}}i_{4} \hspace{0.15cm} \sqrt{3}i_{5} \hspace{0.15cm} \sqrt{2}i_{6})
\end{align*}
This realization satisfies conditions (iv) and (v) of Lemma \ref{thm:Grelssp}, but is neither controllable nor observable.

The aforementioned RLCT networks indicate the importance of removing the assumptions of controllability, observability, and existence of a proper symmetric transfer function from Lemmas \ref{thm:sigsr} and \ref{thm:Grelssp}. This is the objective of this paper. Theorem \ref{thm:ssssr} (resp., \ref{thm:prprop}) generalizes Lemma \ref{thm:sigsr} (resp., \ref{thm:Grelssp}) to systems that need not be controllable. Also, Theorems \ref{thm:sigsymrep} and \ref{thm:prbtp1} extend the results to systems that do not necessarily possess a proper symmetric transfer function. In particular, Theorem \ref{thm:prbtp1} provides necessary and sufficient conditions for a 
behavior to be realizable by an RLCT network, thereby answering the first open problem in \cite{camwb}.

To conclude this section, we discuss some recent developments in the literature on uncontrollable systems, and we contrast these with the results in the present paper. Motivation for developing a theory of reciprocity that does not assume controllability was provided in the behavioral literature in \cite{camwb, JWHVDS}. Indeed, as previously noted, \citet{camwb} stated an open problem that we solve in this paper: \emph{what behaviors are realizable as the port} (driving-point) \emph{behavior of a circuit containing a finite number of passive resistors, capacitors, inductors and transformers}? This question concerns (not necessarily controllable) systems that are both passive and reciprocal. There have since been papers that have considered the question of uncontrollable passive systems \citep[e.g.,][]{THTPLSNA}, and uncontrollable (cyclo)-dissipative systems \citep[e.g.,][]{PBDUS}.\footnote{A system is \emph{cyclo-dissipative} if it has a (not necessarily non-negative) storage function with respect to some supply rate. It is shown in \citep{THAS} that a system is passive (in the sense of Definition \ref{def:pb} of the present paper) if it is cyclo-dissipative with respect to the energy supplied to the system, and the associated storage function is non-negative.} 
But no papers have considered uncontrollable reciprocal systems. For example, consider the behavior $\tilde{\mathcal{B}} := \lbrace (i,v) \in  \mathcal{L}_{1}^{\text{loc}}\left(\mathbb{R}, \mathbb{R}\right) {\times} \mathcal{L}_{1}^{\text{loc}}\left(\mathbb{R}, \mathbb{R}\right) \mid (\tfrac{d}{dt}+1)i = (\tfrac{d}{dt}+1)(\tfrac{dv}{dt}+v)\rbrace$. It has been shown in \citep{THTPLSNA, HUGIFAC} that $\tilde{\mathcal{B}}$ can be realized as the driving-point behavior of an electrical network containing resistors, inductors, transformers \emph{and gyrators} (an RLCTG network). The present paper provides the first proof that (i) this behavior has a signature symmetric realization; and (ii) it can be realized without gyrators (i.e., by an RLCT network).

In fact, as discussed by \citet{JWHVDS}, the subject of uncontrollable reciprocal systems is related to a subtle yet significant question in the development of the theory of uncontrollable (cyclo)-dissipative systems: whether to allow unobservable storage functions. In particular, both \citet{camwb} and \citet{PBDUS} define (cyclo)-dissipativity in terms of the existence of an observable storage function. Yet, in \citet[Section VI]{JWHVDS}, it is demonstrated that systems that are not (cyclo)-dissipative in accordance with this definition can nevertheless possess an unobservable storage function. Moreover, unobservable storage functions arise in electrical networks. In fact, if we consider an uncontrollable behavior with a state-space realization that satisfies the signature symmetry of condition (v) of Lemma \ref{thm:Grelssp}, then it can be shown that this realization is both uncontrollable and unobservable. It can also be shown that any RLCT realization of an uncontrollable behavior necessarily has an unobservable storage function, corresponding to the energy stored in the network's inductors and capacitors.

Given the aforementioned issues with the question of unobservable storage functions, the approach in this paper is aligned with \cite{THTPLSNA}. That paper provided a theory of passivity that does not assume controllability or observability (and also removes other alternative assumptions prevalent in the literature). We refer to that paper for results pertaining to passivity. In this paper, our focus is on developing the theory of (not necessarily controllable) reciprocal systems. 

\section{Reciprocal behaviors}
\label{sec:recip}

Following the motivation outlined in the previous sections, our focus in this paper is on systems of the form:
\begin{align}
\hspace*{-0.3cm} \mathcal{B} &{=} \lbrace (\mathbf{i}, \mathbf{v}) {\in} \mathcal{L}_{1}^{\text{loc}}\left(\mathbb{R}, \mathbb{R}^{n}\right) {\times} \mathcal{L}_{1}^{\text{loc}}\left(\mathbb{R}, \mathbb{R}^{n}\right) \mid P(\tfrac{d}{dt})\mathbf{i} {=} Q(\tfrac{d}{dt})\mathbf{v}\rbrace, \nonumber \\ 
& \text{with } P, Q \in \mathbb{R}^{n \times n}[\xi], \text{normalrank}(\left[P \hspace{0.25cm} {-}Q\right]) = n.\label{eq:bgd}
\end{align}
The driving-point behavior of any passive electrical circuit necessarily has the above form, where $\mathbf{i}$ denotes the driving-point currents and $\mathbf{v}$ the corresponding driving-point voltages \citep[see][]{HUGIFAC}. We note that the partitioning $(\mathbf{i}, \mathbf{v})$ need not be an input-output partition in the sense of \citep[Definition 3.3.1]{JWIMTSC}. Specifically, $Q$ need not be nonsingular, and if $Q$ is nonsingular then $Q^{-1}P$ need not be proper. In this more general setting, it is not possible to define a reciprocal system as a system whose transfer function is symmetric. Instead, we adopt the following definition from \citep[Definition 2.7]{newclms}.
\begin{defn}[Reciprocal system]
\label{def:rb}
Let $\mathcal{B}$ be as in (\ref{eq:bgd}). $\mathcal{B}$ is called \emph{reciprocal} if, for any given $(\mathbf{i}_{a}, \mathbf{v}_{a}), (\mathbf{i}_{b}, \mathbf{v}_{b}) \in \mathcal{B} \cap (\mathcal{D}_{+}\left(\mathbb{R}, \mathbb{R}^{n}\right) \times \mathcal{D}_{+}\left(\mathbb{R}, \mathbb{R}^{n}\right))$, then $\mathbf{v}_{b} \star \mathbf{i}_{a} = \mathbf{i}_{b} \star \mathbf{v}_{a}$.
\end{defn}
\begin{rem}
\textnormal{
Our objective in this paper is to develop a concept of reciprocity that is consistent with the existence of signature symmetric realizations, and the driving-point behaviors of RLCT networks. 
We note that a behavior is reciprocal if and only if its controllable part is reciprocal (this follows from Definition \ref{def:rb} and Lemma \ref{lem:mnir}). In particular, it will follow from Theorems \ref{thm:ssssr} and \ref{thm:prbtp1} that whether a system has a signature symmetric realization depends only on its controllable part, and whether the driving-point behavior of an electric network can be realized without gyrators also depends only on its controllable part.}
\end{rem}

The next theorem shows that any given reciprocal system $\mathcal{B}$ can be transformed into a system of the form of (\ref{eq:bgsqd}) that is also reciprocal (condition \ref{nl:rbtptc4} in Theorem \ref{thm:sigsymrep}). In addition, a necessary and sufficient condition for reciprocity is provided in terms of the polynomial matrices $P$ and $Q$ (condition \ref{nl:rbtptc2} in Theorem \ref{thm:sigsymrep}).

\begin{thm}[Reciprocal behavior theorem, part 1]
\label{thm:sigsymrep}
Let $\mathcal{B}$ be as in (\ref{eq:bgd}). The following are equivalent: 
\begin{enumerate}[label=\arabic*., ref=\arabic*, leftmargin=0.5cm]
\item $\mathcal{B}$ is reciprocal.\label{nl:rbtptc1}
\item $PQ^{T} = QP^{T}$.\label{nl:rbtptc2}
\item There exist real matrices $T_{1} \in \mathbb{R}^{r \times n}$ and $T_{2} \in \mathbb{R}^{(n{-}r) \times n}$ such that (i) $\text{col}(T_{1} \hspace{0.15cm} T_{2})$ is a permutation matrix; and (ii) 
$\hat{\mathcal{B}} := \mathcal{B}^{(\text{col}(T_{1}\mathbf{i} \hspace{0.15cm} {-}T_{2}\mathbf{v}), \text{col}(T_{1}\mathbf{v} \hspace{0.15cm} T_{2}\mathbf{i}))}$ takes the form of (\ref{eq:bgsqd}) and $\hat{Q}^{-1}\hat{P}$ is symmetric.\label{nl:rbtptc4}
\end{enumerate}
\end{thm}

\begin{rem}
\textnormal{
A well known result in behavioral theory is that any behavior $\mathcal{B}$ of the form of (\ref{eq:bgd}) necessarily has an input-output partition. However, condition \ref{nl:rbtptc4} of Theorem \ref{thm:sigsymrep} is not a trivial application of this result. Specifically, in the definition of a reciprocal system (Definition \ref{def:rb}), the system's variables are partitioned into two sets, with an equal number of variables in each set (in the context of electrical networks, these two sets correspond to the driving-point currents and voltages). Condition \ref{nl:rbtptc4} of Theorem \ref{thm:sigsymrep} implies that if the system is reciprocal then it is possible to choose as input a subset of the variables from one of the sets together with the complementary variables from the other set. Note from the example in \citep[Remark 11]{THTPLSNA} that this need not be true if the system is not reciprocal.}
\end{rem} 

We will also show that the system $\hat{\mathcal{B}}$ in condition \ref{nl:rbtptc4} of Theorem \ref{thm:sigsymrep} has a state-space realization $\hat{\mathcal{B}} = \mathcal{B}_{s}^{(\mathbf{u},\mathbf{y})}$ with the properties described in the next theorem.

\begin{thm}[Reciprocal behavior theorem, part 2]
\label{thm:ssssr}
Let $\hat{\mathcal{B}}$ be as in (\ref{eq:bgsqd}). Then the following are equivalent.
\begin{enumerate}[label=\arabic*., ref=\arabic*, leftmargin=0.5cm]
\item $\hat{\mathcal{B}}$ is reciprocal.\label{nl:rbtc1}
\item There exists $\mathcal{B}_{s}$ as in (\ref{eq:bhssr}) and a signature matrix $\Sigma_{i} \in \mathbb{R}^{d \times d}$ such that (i) $\hat{\mathcal{B}} = \mathcal{B}_{s}^{(\mathbf{u},\mathbf{y})}$; and (ii) $A\Sigma_{i} = \Sigma_{i}A^{T}$, $\Sigma_{i}C^{T} = B$, and $D = D^{T}$.\label{nl:rbtc2}
\end{enumerate}
\end{thm}
The two-part reciprocal behavior theorem (Theorems \ref{thm:sigsymrep} and \ref{thm:ssssr}) is proved in Section \ref{sec:rssr}. Then, in Sections \ref{sec:prb}--\ref{sec:raps}, we consider behaviors that are both reciprocal and passive.

\begin{rem}
\label{rem:et}
\textnormal{We emphasise that Lemma \ref{thm:sigsr} is concerned only with controllable systems, whereas Theorem \ref{thm:ssssr} is applicable to any system of the form of (\ref{eq:bgsqd}), irrespective of controllability. 
Note that, if $\hat{\mathcal{B}}$ in (\ref{eq:bgsqd}) is not controllable, and $\mathcal{B}_{s}$ in (\ref{eq:bhssr}) satisfies $\hat{\mathcal{B}} = \mathcal{B}_{s}^{(\mathbf{u},\mathbf{y})}$, then $(A,B)$ cannot be controllable, so Lemma \ref{thm:sigsr} does not apply.} 
\end{rem}

\section{Reciprocity and signature symmetric realizations}
\label{sec:rssr}
The purpose of this section is to prove the reciprocal behavior theorem, parts 1 and 2 (Theorems \ref{thm:sigsymrep} and \ref{thm:ssssr}). We first present the following lemma on the so-called \emph{controllable} and \emph{autonomous} parts of a behavior.
\begin{lem}
\label{lem:mnir}
Let $\mathcal{B}$ be as in (\ref{eq:bgd}). The following hold:
\begin{enumerate}[label=\arabic*., ref=\arabic*, leftmargin=0.5cm]
\item There exist $F, \tilde{P}, \tilde{Q}, U, V \in \mathbb{R}^{n \times n}[\xi]$ such that 
\begin{enumerate}[label=(\roman*), ref=(\roman*)]
\item $P = F\tilde{P}$ and $Q = F\tilde{Q}$; and\label{nl:cadc1i}
\item $\begin{bmatrix}\tilde{P}& \hspace{0.1cm}{-}\tilde{Q}\\ U& \hspace{0.1cm}V\end{bmatrix}$ is unimodular.\label{nl:cadc1ii}
\end{enumerate}
Also, if $F, \tilde{P}, \tilde{Q}, U, V \in \mathbb{R}^{n \times n}[\xi]$ satisfy \ref{nl:cadc1i}--\ref{nl:cadc1ii}; \newline $\mathcal{B}_{c} := \lbrace (\mathbf{i}, \mathbf{v}) \in \mathcal{L}_{1}^{\text{loc}}\left(\mathbb{R}, \mathbb{R}^{n}\right) \times \mathcal{L}_{1}^{\text{loc}}\left(\mathbb{R}, \mathbb{R}^{n}\right) \mid \tilde{P}(\tfrac{d}{dt})\mathbf{i} = \tilde{Q}(\tfrac{d}{dt})\mathbf{v} \rbrace$; and \newline $\mathcal{B}_{a} := \lbrace (\mathbf{i}, \mathbf{v}) \in \mathcal{L}_{1}^{\text{loc}}\left(\mathbb{R}, \mathbb{R}^{n}\right) \times \mathcal{L}_{1}^{\text{loc}}\left(\mathbb{R}, \mathbb{R}^{n}\right) \mid P(\tfrac{d}{dt})\mathbf{i} = Q(\tfrac{d}{dt})\mathbf{v} \text{ and } U(\tfrac{d}{dt})\mathbf{i} = -V(\tfrac{d}{dt})\mathbf{v} \rbrace$, \newline then 
$(\mathbf{i}, \mathbf{v}) \in \mathcal{B} \iff$ there exist $(\mathbf{i}_{1}, \mathbf{v}_{1}) \in \mathcal{B}_{c}$ and $(\mathbf{i}_{2}, \mathbf{v}_{2}) \in \mathcal{B}_{a}$ with $\mathbf{i} = \mathbf{i}_{1} {+} \mathbf{i}_{2}$ and $\mathbf{v} = \mathbf{v}_{1} {+} \mathbf{v}_{2}$.\label{nl:cadc1}
\item There exist $M, N \in \mathbb{R}^{n \times n}[\xi]$ such that
\begin{enumerate}[label=(\roman*), ref=(\roman*)]
\item $PM = QN$; and\label{nl:cadc2i}
\item $\text{rank}(\text{col}(M \hspace{0.15cm} N)(\lambda)) = n$ for all $\lambda \in \mathbb{C}$.\label{nl:cadc2ii}
\end{enumerate}
Also, if $M, N \in \mathbb{R}^{n \times n}[\xi]$ satisfy \ref{nl:cadc2i}--\ref{nl:cadc2ii}, then $(\mathbf{i}, \mathbf{v}) \in \mathcal{B} \cap (\mathcal{D}_{+}\left(\mathbb{R}, \mathbb{R}^{n}\right) \times \mathcal{D}_{+}\left(\mathbb{R}, \mathbb{R}^{n}\right))$ if and only if there exists $\mathbf{z} \in \mathcal{D}_{+}\left(\mathbb{R}, \mathbb{R}^{n}\right)$ such that $\mathbf{i} = M(\tfrac{d}{dt})\mathbf{z}$ and $\mathbf{v} = N(\tfrac{d}{dt})\mathbf{z}$. In particular, $(\mathbf{i}, \mathbf{v}) \in \mathcal{B}_{c}$.
\end{enumerate}
\end{lem}
\begin{pf}
This requires only minor modifications to the proof of Lemma 17 in \citep{THTPLSNA}.
\end{pf}

{\bf PROOF OF THEOREM \ref{thm:sigsymrep}}  (see p.\ \pageref{thm:sigsymrep}). We let $M$ and $N$ be as in Lemma \ref{lem:mnir}, and we will show the equivalence of conditions \ref{nl:rbtptc1}--\ref{nl:rbtptc4} to the additional condition:
\begin{enumerate}[label=\arabic*., ref=\arabic*, leftmargin=0.5cm]
\setcounter{enumi}{3}
\item $M^{T}N = N^{T}M$.\label{nl:rbtptc3}
\end{enumerate}
Specifically, we will prove \ref{nl:rbtptc1} $\iff$ \ref{nl:rbtptc3} $\iff$ \ref{nl:rbtptc4} $\iff$ \ref{nl:rbtptc2}.

{\bf \ref{nl:rbtptc1} $\boldsymbol{\iff}$ \ref{nl:rbtptc3}.} \hspace{0.3cm} 
From Lemma \ref{lem:mnir}, there exist $\mathbf{z}_{1}, \mathbf{z}_{2} \in \mathcal{D}_{+}\left(\mathbb{R}, \mathbb{R}^{n}\right)$ such that 
\begin{equation*}
\hspace*{-0.2cm} \mathbf{i}_{1} {=} M(\tfrac{d}{dt})\mathbf{z}_{1}, \mathbf{i}_{2} {=} M(\tfrac{d}{dt})\mathbf{z}_{2}, \mathbf{v}_{1} {=} N(\tfrac{d}{dt})\mathbf{z}_{1} \text{ and } \mathbf{v}_{2} {=} N(\tfrac{d}{dt})\mathbf{z}_{2}.
\end{equation*}
Now, consider a fixed but arbitrary $t_{0} \in \mathbb{R}$, and let 
\begin{equation*}
\hat{\mathbf{z}}_{2}(t) = \mathbf{z}_{2}(t_{0}-t) \text{ for all } t \in \mathbb{R}.
\end{equation*}
Then $\hat{\mathbf{z}}_{2} \in \mathcal{D}_{-}\left(\mathbb{R}, \mathbb{R}^{n}\right)$, $\mathbf{i}_{2}(t_{0}-\tau) = (M(-\tfrac{d}{dt})\hat{\mathbf{z}}_{2})(\tau)$ and $\mathbf{v}_{2}(t_{0}-\tau) = (N(-\tfrac{d}{dt})\hat{\mathbf{z}}_{2})(\tau)$ for all $\tau \in \mathbb{R}$. Thus,
\begin{align*}
(\mathbf{v}_{2} \star \mathbf{i}_{1})(t_{0}) &= \smallint_{-\infty}^{\infty} (N(-\tfrac{d}{dt})\hat{\mathbf{z}}_{2})(\tau)^{T}(M(\tfrac{d}{dt})\mathbf{z}_{1})(\tau)d\tau, \text{ and} \\
(\mathbf{i}_{2} \star \mathbf{v}_{1})(t_{0}) &=  \smallint_{-\infty}^{\infty}(M(-\tfrac{d}{dt})\hat{\mathbf{z}}_{2})(\tau)^{T}(N(\tfrac{d}{dt})\mathbf{z}_{1})(\tau)d\tau .
\end{align*}
It follows from  \citep[Section 2.2]{AVDSSMIP} that 
\begin{equation*}
\hspace*{-0.3cm}(\mathbf{v}_{2} \star \mathbf{i}_{1} {-} \mathbf{i}_{2} \star \mathbf{v}_{1})(t_{0}) = \smallint_{-\infty}^{\infty}\hat{\mathbf{z}}_{2}(\tau)^{T} ((N^{T}M {-} M^{T}N)(\tfrac{d}{dt})\mathbf{z}_{1})(\tau)d\tau .
\end{equation*}
Since $t_{0}$ is arbitrary, then we conclude that 
$\mathcal{B}$ is reciprocal if and only if the above integral is zero for all $\mathbf{z}_{1} \in \mathcal{D}_{+}\left(\mathbb{R}, \mathbb{R}^{n}\right)$ and $\hat{\mathbf{z}}_{2} \in \mathcal{D}_{-}\left(\mathbb{R}, \mathbb{R}^{n}\right)$. In particular, if $N^{T}M = M^{T}N$, then $\mathcal{B}$ is reciprocal. Conversely, note that if the above integral is zero for all $\mathbf{z}_{1} \in \mathcal{D}_{+}\left(\mathbb{R}, \mathbb{R}^{n}\right)$ and $\hat{\mathbf{z}}_{2} \in \mathcal{D}_{-}\left(\mathbb{R}, \mathbb{R}^{n}\right)$, then $(N^{T}M - M^{T}N)(\tfrac{d}{dt})\mathbf{z}_{1} \equiv 0$ for all $\mathbf{z}_{1} \in \mathcal{D}\left(\mathbb{R}, \mathbb{R}^{n}\right)$ (since otherwise the integral is strictly positive with $\hat{\mathbf{z}}_{2} = (N^{T}M - M^{T}N)(\tfrac{d}{dt})\mathbf{z}_{1}$). It then follows from \citep[Sections 2.5.6 and 3.3]{JWIMTSC} that $N^{T}M = M^{T}N$.

{\bf \ref{nl:rbtptc3} $\boldsymbol{\Rightarrow}$ \ref{nl:rbtptc4}.} \hspace{0.3cm} First, bring $\text{col}(M \hspace{0.15cm} N)$ into column proper form. In other words, let $U$ be a unimodular matrix with
\begin{equation*}
\begin{bmatrix}M \\ N\end{bmatrix}U =: W,
\end{equation*}
in which the leading coefficient matrix $W^{L}$ of $W$ has full column rank \citep[see][Section 2.5]{WLMS}. Next, partition $W^{L}$ compatibly with $\text{col}(M \hspace{0.15cm} N)$ as $W^{L} = \text{col}(W_{1}^{L} \hspace{0.15cm} W_{2}^{L})$, let $r$ denote the rank of $W_{1}^{L}$, permute the columns of $W_{1}^{L}$ so the first $r$ columns are linearly independent, and then permute the rows so the first $r$ rows are linearly independent. This gives permutation matrices $T = \text{col}(T_{1} \hspace{0.15cm} T_{2}) \in \mathbb{R}^{n \times n}$ and $S = [S_{1} \hspace{0.15cm} S_{2}] \in \mathbb{R}^{n \times n}$ and an $X \in \mathbb{R}^{r \times (n{-}r)}$ such that
\begin{equation*}
\begin{bmatrix}\hat{M}\\ \hat{N}\end{bmatrix} = \begin{bmatrix}\hat{M}_{11}& \hat{M}_{12}\\ \hat{M}_{21}& \hat{M}_{22}\\ \hline \hat{N}_{11}& \hat{N}_{12}\\ \hat{N}_{21}& \hat{N}_{22}\end{bmatrix} := \begin{bmatrix}T_{1}& 0\\0& -T_{2}\\ \hline 0&T_{1}\\ T_{2}& 0\end{bmatrix}\begin{bmatrix}M\\ N\end{bmatrix}U\begin{bmatrix}S_{1}& S_{2}\end{bmatrix}
\end{equation*}
is in column proper form, and its leading coefficient matrix $\text{col}(\hat{M}^{L} \hspace{0.15cm} \hat{N}^{L})$ takes the form
\begin{equation*}
\begin{bmatrix}\hat{M}^{L}\\ \hat{N}^{L}\end{bmatrix} = 
\begin{bmatrix}\hat{M}_{11}^{L}& \hat{M}_{11}^{L}X\\ \hat{M}_{21}^{L}& \hat{M}_{22}^{L}\\ \hline \hat{N}_{11}^{L}& \hat{N}_{12}^{L}\\ \hat{N}_{21}^{L}& \hat{N}_{21}^{L}X\end{bmatrix} = \begin{bmatrix}T_{1}& 0\\0& -T_{2}\\ \hline 0&T_{1}\\ T_{2}& 0\end{bmatrix}\begin{bmatrix}W_{1}^{L}\\ W_{2}^{L}\end{bmatrix}\begin{bmatrix}S_{1}& S_{2}\end{bmatrix}
\end{equation*}
where $\hat{M}_{11}^{L}$ is nonsingular. It is then easily verified that $\hat{M}^{T}\hat{N} - \hat{N}^{T}\hat{M} = (US)^{T}(M^{T}N - N^{T}M)(US) = 0$. We will show that $\hat{M}^{L}$ is nonsingular, and it follows that $\hat{N}\hat{M}^{-1}$ is proper \citep[see][Section 2]{PRSMLS}. We then let $\hat{P} := [PT_{1}^{T} \hspace{0.2cm} QT_{2}^{T}]$ and $\hat{Q} := [QT_{1}^{T} \hspace{0.25cm} {-}PT_{2}^{T}]$, we recall that $PM = QN$, and it is then easily verified that $\hat{P}\hat{M} = \hat{Q}\hat{N}$. This implies that $\hat{Q}$ is nonsingular with $\hat{Q}^{-1}\hat{P} = \hat{N}\hat{M}^{-1}$, which is symmetric since $\hat{N}\hat{M}^{-1} = (\hat{M}^{-1})^{T}\hat{M}^{T}\hat{N}\hat{M}^{-1} = (\hat{M}^{-1})^{T}\hat{N}^{T}\hat{M}\hat{M}^{-1} = (\hat{M}^{-1})^{T}\hat{N}^{T}$. Finally, with $\mathbf{i}_{1} := T_{1}\mathbf{i}, \mathbf{v}_{1} := T_{1}\mathbf{v}, \mathbf{i}_{2} := T_{2}\mathbf{i}$ and $\mathbf{v}_{2} := T_{2}\mathbf{v}$, then it is easily shown that $\hat{\mathcal{B}}$ takes the form indicated in the present theorem statement.

To complete the proof of the present implication, it remains to show that if $\mathbf{z} \in \mathbb{R}^{n}$ and $\hat{M}^{L}\mathbf{z} = 0$ then $\mathbf{z} = 0$. To see this, we denote the column degree of the $j$th column of $\text{col}(\hat{M} \hspace{0,15cm} \hat{N})$ by $\hat{d}_{j}$, and we note that the entry in the $i$th row and $j$th column of $(\hat{M}^{L})^{T}\hat{N}^{L} - (\hat{N}^{L})^{T}\hat{M}^{L}$ is the coefficient of $\xi^{\hat{d}_{i} + \hat{d}_{j}}$ in the entry in the $i$th row and $j$th column of $\hat{M}^{T}\hat{N} - \hat{N}^{T}\hat{M}$, which is necessarily zero. Now, let $\mathbf{z} \in \mathbb{R}^{n}$ satisfy $\hat{M}^{L}\mathbf{z} = 0$. Then $\hat{M}_{11}^{L}[I \hspace{0.15cm} X]\mathbf{z} = 0$. Since $\hat{M}_{11}^{L}$ is nonsingular, it follows that there exists $\mathbf{w} \in \mathbb{R}^{n{-}r}$ such that $\mathbf{z} = \text{col}({-}X \hspace{0.15cm} I)\mathbf{w}$. But 
%
\begin{align*}
0 &= \begin{bmatrix}I& 0\end{bmatrix}((\hat{M}^{L})^{T}\hat{N}^{L} - (\hat{N}^{L})^{T}\hat{M}^{L})\begin{bmatrix}-X\\ I\end{bmatrix}\mathbf{w} \\
&= (\hat{M}_{11}^{L})^{T}\begin{bmatrix}\hat{N}_{11}^{L}& \hat{N}_{12}^{L}\end{bmatrix}\begin{bmatrix}-X\\ I\end{bmatrix}\mathbf{w}.
\end{align*}
Since $\hat{M}_{11}^{L}$ is nonsingular, then $[\hat{N}_{11}^{L} \hspace{0.15cm} \hat{N}_{12}^{L}]\mathbf{z} = 0$. It follows that $\text{col}(\hat{M}^{L} \hspace{0.15cm} \hat{N}^{L})\mathbf{z} = 0$. But $\text{col}(\hat{M}^{L} \hspace{0.15cm} \hat{N}^{L})$ has full column rank as $\text{col}(\hat{M} \hspace{0.15cm} \hat{N})$ is in column proper form, and we conclude that $\mathbf{z} = 0$.

{\bf \ref{nl:rbtptc4} $\boldsymbol{\Rightarrow}$ \ref{nl:rbtptc3}.} \hspace{0.3cm} Let 
$\hat{M}, \hat{N} \in \mathbb{R}^{n \times n}[\xi]$ be such that the columns of $\text{col}(\hat{M} \hspace{0.15cm} \hat{N})$ are a basis for the right syzygy of $[\hat{P} \hspace{0.25cm} {-}\hat{Q}]$ \citep[see][p.\ 85]{JWBAOIS}. Similar to before, we find that $\hat{M}$ is nonsingular and $\hat{N}\hat{M}^{-1} = \hat{Q}^{-1}\hat{P}$, which is symmetric. Also, there exists a unimodular $U$ such that
\begin{equation*}
\begin{bmatrix}M\\ N\end{bmatrix}U = \begin{bmatrix}T_{1}^{T}& 0& 0& T_{2}^{T}\\ 0& -T_{2}^{T}& T_{1}^{T}& 0\end{bmatrix}\begin{bmatrix}\hat{M}\\ \hat{N}\end{bmatrix}.
\end{equation*}
This follows from \citep[][pp.\ 84--85]{JWBAOIS}, noting from the definition of $\mathcal{B}$ and $\hat{\mathcal{B}}$ that the columns of the matrix on the right hand side of the above equation span the right syzygy of $[P \hspace{0.25cm} {-}Q]$. It can then be verified that $U^{T}(M^{T}N - N^{T}M)U = \hat{M}^{T}\hat{N} - \hat{N}^{T}\hat{M} = 0$. Since $U$ is nonsingular, this implies that $M^{T}N - N^{T}M = 0$.

{\bf \ref{nl:rbtptc4} $\boldsymbol{\iff}$ \ref{nl:rbtptc2}.} \hspace{0.3cm} The proof is analogous to \ref{nl:rbtptc3} $\iff$ \ref{nl:rbtptc4}. \qed

{\bf PROOF OF THEOREM \ref{thm:ssssr}} (see p.\ \pageref{thm:ssssr}). That \ref{nl:rbtc2} $\Rightarrow$ \ref{nl:rbtc1} follows from Theorem \ref{thm:sigsymrep}, noting from the proof of Lemma \ref{thm:sigsr} that $(\hat{Q}^{-1}\hat{P})(\xi) = D + C(\xi I - A)^{-1}B$, which is symmetric. To see that \ref{nl:rbtc1} $\Rightarrow$ \ref{nl:rbtc2}, note initially that if $\hat{\mathcal{B}}$ is controllable then the result follows from Lemma \ref{thm:sigsr} and Theorem \ref{thm:sigsymrep}. 
Otherwise, following \citep[Notes A.1 and A.3]{THAS} and \citep[Corollary 5.2.25]{JWIMTSC}, we can construct a realization $(\tilde{A}, \tilde{B}, \tilde{C}, D)$ of $(\hat{P}, \hat{Q})$ such that $(\tilde{C},\tilde{A})$ is observable; and 
\begin{equation*}
\tilde{A} = \begin{bmatrix}\tilde{A}_{11}& \tilde{A}_{12}\\ 0 & \tilde{A}_{22}\end{bmatrix}, \hspace{0.1cm} \tilde{B} = \begin{bmatrix}\tilde{B}_{1}\\ 0\end{bmatrix}, \hspace{0.1cm} \tilde{C} = \begin{bmatrix}\tilde{C}_{1}& \tilde{C}_{2}\end{bmatrix},
\end{equation*}
where $(\tilde{A}_{11}, \tilde{B}_{1})$ is controllable, $(\tilde{C}_{1}, \tilde{A}_{11})$ is observable, and $\hat{Q}^{-1}\hat{P}(\xi) = \tilde{C}_{1}(\xi I - \tilde{A}_{11})^{-1}\tilde{B}_{1} + D$, which is symmetric. 
It then follows from the proof of Lemma \ref{thm:sigsr} that there exists a symmetric $P$ such that $P\tilde{A}_{11} {=} \tilde{A}_{11}^{T}P$, $\tilde{C}_{1}^{T} {=} P\tilde{B}_{1}$, and $D {=} D^{T}$. Now, let
\begin{align*}
\hat{A} := \begin{bmatrix}\tilde{A}_{11}& \tilde{A}_{12}& 0\\ 0 & \tilde{A}_{22}& 0\\ \tilde{A}_{12}^{T}P& 0& \tilde{A}_{22}^{T}\end{bmatrix}, \hspace{0.1cm} \hat{C} := \begin{bmatrix}\tilde{C}_{1}& \tilde{C}_{2} & 0\end{bmatrix}, \\
\hat{B} := \begin{bmatrix}\tilde{B}_{1}\\ 0\\ \tilde{C}_{2}^{T}\end{bmatrix}, \text{ and } S := \begin{bmatrix}P& 0& 0\\ 0 & 0& I\\ 0& I& 0\end{bmatrix}.
\end{align*}  
Since $(\tilde{A}, \tilde{B}, \tilde{C}, D)$ is a realization for $(\hat{P}, \hat{Q})$, then so too is $(\hat{A}, \hat{B}, \hat{C}, D)$ (this follows from Remark \ref{rem:ssr}, as $\hat{C}\hat{A}^{k} = [\tilde{C}\tilde{A}^{k} \hspace{0.15cm} 0]$ for $k = 0, 1, 2, \ldots$). 
Also, $S\hat{A} = \hat{A}^{T}S$ and $S\hat{B} = \hat{C}^{T}$. Finally, as $P$ is symmetric, there exists a real matrix $R$ and a signature matrix $\tilde{\Sigma}_{i}$ such that $P = R^{T}\tilde{\Sigma}_{i}R$, and we define $T$ and $\Sigma_{i}$ (partitioned compatibly) as
\begin{equation*}
T:= \begin{bmatrix}R& 0& 0\\ 0& \tfrac{1}{\sqrt{2}}I& -\tfrac{1}{\sqrt{2}}I\\ 0& \tfrac{1}{\sqrt{2}}I& \tfrac{1}{\sqrt{2}}I\end{bmatrix}, \text{ and } \Sigma_{i} := \begin{bmatrix}\tilde{\Sigma}_{i}& 0& 0\\ 0& -I& 0\\ 0& 0& I\end{bmatrix}.
\end{equation*}
Then $S = T^{T}\Sigma_{i}T$, and $A := T\hat{A}T^{-1}$, $B:= T\hat{B}$, $C := \hat{C}T^{-1}$ satisfy the conditions of the present theorem. \qed

\begin{rem}
\label{rem:compsm}
\textnormal{
Note that the proofs of Lemma \ref{thm:sigsr} and Theorem \ref{thm:ssssr} provide an algorithm for the construction of the realization $(A,B,C,D)$ and the signature matrix $\Sigma_{i}$ in Theorem \ref{thm:ssssr}. Specifically, $P$ in the proof of that theorem can be obtained from the explicit formula in the proof of Lemma \ref{thm:sigsr}, whereupon $R$ and $\tilde{\Sigma}_{i}$ can be obtained from an eigenvalue decomposition for $P$.
}
\end{rem}

\section{Passive and reciprocal behaviors}
\label{sec:prb}

In this section, we present our main results concerning passive and reciprocal systems. We define passivity in accordance with \citep[][Definition 5]{THTPLSNA} as follows.
\begin{defn}[Passive system]
\label{def:pb}
$\mathcal{B}$ in (\ref{eq:bgd}) is called \emph{passive} if, given any $(\mathbf{i}, \mathbf{v}) \in \mathcal{B}$ and any $t_{0} \in \mathbb{R}$, there exists a $K \in \mathbb{R}$ (dependent on $(\mathbf{i}, \mathbf{v})$ and $t_{0}$) such that, if $t_{1} \geq t_{0}$ and $(\tilde{\mathbf{i}}, \tilde{\mathbf{v}}) \in \mathcal{B}$ satisfies $(\tilde{\mathbf{i}}(t), \tilde{\mathbf{v}}(t)) = (\mathbf{i}(t), \mathbf{v}(t))$ for $t < t_{0}$, then $-\smallint_{t_{0}}^{t_{1}} \tilde{\mathbf{i}}^{T}(t)\tilde{\mathbf{v}}(t) dt < K$.
\end{defn}
\begin{rem}
\textnormal{Here, $-\smallint_{t_{0}}^{t_{1}} \tilde{\mathbf{i}}^{T}(t)\tilde{\mathbf{v}}(t) dt$ is the net energy supplied to the system between $t_{0}$ and $t_{1}$, and the bound $K$ is necessarily non-negative (as the integral is zero when $t_{1} = t_{0}$). Thus, Definition \ref{def:pb} formalises the concept that a system is passive if the net energy that can be extracted from the system into the future is bounded above (this bound depending only on the past trajectory of the system). It is shown in \cite{THAS} that this definition is consistent with the existence of a non-negative quadratic state storage function with respect to the energy supplied to the system.}
\end{rem}

The following concept of a positive-real pair was introduced by \citet{THTPLSNA}, where it was shown that $\mathcal{B}$ in (\ref{eq:bgd}) is passive if and only if $(P,Q)$ is a positive-real pair. 
\begin{defn}[Positive-real pair]
\label{def:prp}
Let $P, Q \in \mathbb{R}^{n \times n}[\xi]$. 
We call $(P, Q)$ a \emph{positive-real pair} if:
\begin{enumerate}[label=(\alph*)]
\item $P(\lambda)Q(\bar{\lambda})^{T} + Q(\lambda)P(\bar{\lambda})^{T} \geq 0$ for all $\lambda \in \mathbb{C}_{+}$; \label{nl:prpc1}
\item $\text{rank}([P \hspace{0.25cm} {-}Q](\lambda)) = n$ for all $\lambda \in \overbar{\mathbb{C}}_{+}$; and \label{nl:prpc2} 
\item if $\mathbf{p} \in \mathbb{R}^{n}[\xi]$ and $\lambda \in \mathbb{C}$ satisfy $\mathbf{p}(\xi)^{T}(P(\xi)Q(-\xi)^{T} + Q(\xi)P(-\xi)^{T}) = 0$ and $\mathbf{p}(\lambda)^{T}[P \hspace{0.25cm} {-}Q](\lambda) = 0$, then $\mathbf{p}(\lambda) = 0$. \label{nl:prpc3}
\end{enumerate}
\end{defn}

\begin{rem}
\textnormal{
Note that, in contrast with reciprocity, it is possible for the controllable part of a system to be passive yet for the system itself to not be passive. E.g., let $\mathcal{B}_{s}$ be as in (\ref{eq:bhssr}) with $B = 0$, $C = 1$ and $D = 1$, so $D + C(\xi I - A)^{-1}B = 1$ for all $A \in \mathbb{R}$ (i.e., the controllable part of the system is independent of $A$). From Remark \ref{rem:ssr}, if $A = -1$, then for any given $(u,y) \in \mathcal{B}_{s}^{(u,y)}$, there exists $k_{1} \in \mathbb{R}$ such that $-\smallint_{t_{0}}^{t_{1}}(uy)(t)dt = \smallint_{t_{0}}^{t_{1}}-(u(t)+\tfrac{k_{1}}{2}e^{-t})^{2} + \tfrac{k_{1}^{2}}{4}e^{-2t}dt \leq \tfrac{k_{1}^{2}}{8}e^{-2t_{0}}$, so this system is passive. But if $A = 0$, then there exists $(u,y) \in \mathcal{B}_{s}^{(u,y)}$ with $u = -\tfrac{k_{2}}{2} = -y$ for all $t \in \mathbb{R}$, in which case $-\smallint_{t_{0}}^{t_{1}}(uy)(t)dt = \tfrac{k_{2}^{2}}{4}(t_{1}-t_{0})$, so this system is not passive.}
\end{rem}

In the following theorem, we state necessary and sufficient algebraic conditions for $\mathcal{B}$ in (\ref{eq:bgd}) to be passive and reciprocal. We also show that these conditions are equivalent to $\mathcal{B}$ being realizable by an RLCT network, thus solving the first open problem in \cite{camwb}.
\begin{thm}\textnormal{{\bf (Passive and reciprocal behavior theorem, part 1)}}
\label{thm:prbtp1}
Let $\mathcal{B}$ be as in (\ref{eq:bgd}). The following are equivalent:
\begin{enumerate}[label=\arabic*., ref=\arabic*, leftmargin=0.5cm]
\item $\mathcal{B}$ is passive and reciprocal.\label{nl:prbtp1c1}
\item $(P,Q)$ is a positive-real pair and $PQ^{T} = QP^{T}$.\label{nl:prbtp1c2}
\item $\mathcal{B}$ is the driving-point behavior of an RLCT network.\label{nl:prbtp1c3}
\end{enumerate}
\end{thm}

In our final theorem, we generalize Lemma \ref{thm:Grelssp} to systems that need not be controllable.
\begin{thm}\textnormal{{\bf (Passive and reciprocal behavior theorem, part 2)}}
\label{thm:prprop}
Let $\hat{\mathcal{B}}$ be as in (\ref{eq:bgsqd}). Then the following are equivalent.
\begin{enumerate}[label=\arabic*., ref=\arabic*, leftmargin=0.5cm]
\item $\hat{\mathcal{B}}$ is passive and reciprocal.\label{nl:prbtc1}
\item There exists $\mathcal{B}_{s}$ as in (\ref{eq:bhssr}) and a signature matrix $\Sigma_{i} \in \mathbb{R}^{d \times d}$ such that
\begin{enumerate}[label=(\roman*), ref=(\roman*), leftmargin=0.5cm]
\item $\hat{\mathcal{B}} = \mathcal{B}_{s}^{(\mathbf{u},\mathbf{y})}$;
\item $\begin{bmatrix}{-}A& {-}B\\ C& D\end{bmatrix}\! {+} \!\begin{bmatrix}{-}A& {-}B\\ C& D\end{bmatrix}^{T} \geq 0$; and
\item $\Sigma_{i}A = A^{T}\Sigma_{i}$, $\Sigma_{i}B = C^{T}$, and $D = D^{T}$.
\end{enumerate}\label{nl:prbtc2}
\end{enumerate}
\end{thm}
The two-part passive and reciprocal behavior theorem (Theorems \ref{thm:prbtp1} and \ref{thm:prprop}) is proved in Section \ref{sec:raps}. The proofs can be combined with existing results in the literature to obtain a passive and reciprocal realization for any given passive and reciprocal system of the form of (\ref{eq:bgsqd}), and to obtain an RLCT realization for an arbitrary given passive and reciprocal system of the form of (\ref{eq:bgd}). This will be illustrated by two examples in Section \ref{sec:ex}. 

\section{Proof of the passive and reciprocal behavior theorem}
\label{sec:raps}

The purpose of this section is to prove Theorems \ref{thm:prbtp1} and \ref{thm:prprop}. These two theorems will be proved in reverse order. First, we prove the following result, which uses the supplementary lemmas in Appendix \ref{sec:prbtsl}. 

\begin{lem}
\label{thm:ppst}
Let $\hat{\mathcal{B}}$ in (\ref{eq:bgsqd}) be passive and reciprocal. Then there exists $\mathcal{B}_{s}$ as in (\ref{eq:bhssr}) such that $\hat{\mathcal{B}} = \mathcal{B}_{s}^{(\mathbf{u},\mathbf{y})}$ and the following properties both hold:
\begin{enumerate}[label=\arabic*., ref=\arabic*, leftmargin=0.5cm]
\item there exists $X \in \mathbb{R}^{d \times d}$ such that $X > 0$ and $\begin{bmatrix} -XA - A^{T}X& C^{T} - XB\\ C - B^{T}X& D + D^{T}\end{bmatrix} \geq 0$; and \label{nl:prc1}
\item there exists a symmetric nonsingular $S \in \mathbb{R}^{d \times d}$ such that $SA = A^{T}S$, $SB = C^{T}$, and $D = D^{T}$.\label{nl:prc2}
\end{enumerate}
\end{lem}

\begin{pf}
We will prove this first for the case in which $D + D^{T} > 0$, and then for the general case.

{\bf Case (i): $\boldsymbol{D + D^{T} > 0}$.} \hspace{0.3cm} We let $\hat{A}, \hat{B}, \hat{C}, D$ and $S$ be as in the proof of Theorem \ref{thm:ssssr}, and we let $A = \hat{A}, B = \hat{B}$ and $C = \hat{C}$. From that proof, condition \ref{nl:prc2} of the present theorem statement holds. Also, $\hat{\mathcal{B}} = \mathcal{B}_{s}^{(\mathbf{u},\mathbf{y})}$ is passive and $(C,A)$ is detectable. Thus, from Lemma \ref{lem:pdrr}, there exists $K \in \mathbb{R}^{d \times d}$ such that $K > 0$ and $\Upsilon(K) \geq 0$, where $\Upsilon(K)$ is as in (\ref{eq:upapsid}). It can then be verified that $X := K^{-1}$ satisfies condition \ref{nl:prc1} of the present theorem statement. 

{\bf Case (ii): general case.} \hspace{0.3cm} Let $P_{1} := \hat{P}$ and $Q_{1} := \hat{Q}$, and consider the following three statements \citep[c.f.,][proof of Theorem 13]{THAS}:
\begin{remunerate}
\labitem{(R\arabic{muni})}{nl:ip1} $P_{i}, Q_{i} \in \mathbb{R}^{n_{i} \times n_{i}}[\xi]$ where $(P_{i}, Q_{i})$ is a positive-real pair and $Q_{i}^{-1}P_{i}$ is proper and symmetric.
\labitem{(R\arabic{muni})}{nl:ip3} $P_{i}, Q_{i}$ are as in \ref{nl:ip1}, $P_{i}$ is nonsingular, and $\lim_{\xi \rightarrow \infty}((Q_{i}^{-1}P_{i})(\xi)) = \text{diag}(I_{r_{i}} \hspace{0.15cm} 0)$.
\labitem{(R\arabic{muni})}{nl:ip4} $P_{i}, Q_{i}$ are as in \ref{nl:ip1}, and either $n_{i} = 0$ or $\lim_{\xi \rightarrow \infty}((Q_{i}^{-1}P_{i})(\xi)) = I$.
\end{remunerate}
By \cite[Theorem 7]{THTPLSNA} and Theorem \ref{thm:sigsymrep} of this paper, $P_{1}, Q_{1}$ satisfy condition \ref{nl:ip1}. Then, using Lemmas \ref{lem:gprl2} and \ref{lem:lrz} (see Appendix \ref{sec:prbtsl}), we construct $P_{2}, \ldots, P_{m}$, $Q_{2}, \ldots, Q_{m}$ such that condition \ref{nl:ip1} is satisfied, $n_{i} \leq n_{i-1}$, and $\deg{(\det{(Q_{i})})} \leq \deg{(\det{(Q_{i-1})})}$, for $i = 2, \ldots , m$; and
\begin{enumerate}[leftmargin=0.5cm]
\item if, for $i = k-1$, \ref{nl:ip3} is not satisfied, then \ref{nl:ip3} is satisfied for $i = k$, and if $P_{k{-}1}$ is singular then $n_{k} < n_{k{-}1}$ (Lemma \ref{lem:gprl2}); and
\item if, for $i = k{-}1$, \ref{nl:ip3} is satisfied but \ref{nl:ip4} is not, then $\deg{(\det{(Q_{k})})} < \deg{(\det{(Q_{k{-}1})})}$ (Lemma \ref{lem:lrz}).
\end{enumerate}
This inductive procedure terminates in a finite number of steps with polynomial matrices $P_{m}$ and $Q_{m}$ that satisfy conditions \ref{nl:ip1}--\ref{nl:ip4}. An example is given in Section \ref{sec:ex}.

Next, we consider the following three statements:
\begin{remunerate}
\labitem{(S\arabic{muni})}{nl:csn1} There exist polynomial matrices $\mathcal{A}_{i}(\xi) {:=} \xi I - A_{i}$, $Y_{i}, Z_{i}, U_{i}, V_{i}, E_{i}, F_{i}$, and $G_{i}$, with $G_{i}$ nonsingular, and \newline
$\begin{bmatrix}Y_{i}& Z_{i} \\ U_{i}& V_{i}\end{bmatrix}\begin{bmatrix}-D_{i}& I& -C_{i}\\ -B_{i}& 0& \mathcal{A}_{i}\end{bmatrix} = \begin{bmatrix}-P_{i}& Q_{i}& 0\\ -E_{i}& -F_{i}& G_{i}\end{bmatrix}$,\newline
where the leftmost matrix is unimodular.
\labitem{(S\arabic{muni})}{nl:csn2-} The matrix $X_{i} \in \mathbb{R}^{d_{i} \times d_{i}}$ satisfies $X_{i}> 0$ and $\Omega_{i}(X_{i}) := \!\begin{bmatrix}-A_{i}^{T}X_{i} {-} X_{i}A_{i}& C_{i}^{T}{-}X_{i}B_{i}\\ C_{i}{-}B_{i}^{T}X_{i}& D_{i} {+} D_{i}^{T}\end{bmatrix}\! \geq 0$.

\labitem{(S\arabic{muni})}{nl:csn2} There exists a symmetric $S_{i} \in \mathbb{R}^{d_{i} \times d_{i}}$ such that $S_{i}A_{i} = A_{i}^{T}S_{i}$, $S_{i}B_{i} = C_{i}^{T}$ and $D_{i} = D_{i}^{T}$.
\end{remunerate}
From case (i) and Lemma \ref{lem:ssr}, there exist real matrices $A_{m}, B_{m}, C_{m}, D_{m}, X_{m}$ and $S_{m}$ such that \ref{nl:csn1}--\ref{nl:csn2} hold for $i = m$. Then, using Lemmas \ref{lem:gprl2} and \ref{lem:lrz}, we find that there exist real matrices $A_{i}, B_{i}, C_{i}, D_{i}, X_{i}$ and $S_{i}$ such that \ref{nl:csn1}--\ref{nl:csn2} hold for $i = m-1, \ldots, 1$. Since $P = P_{1}$ and $Q = Q_{1}$, then letting $A = A_{1}, B = B_{1}, C = C_{1}, D = D_{1}, S = S_{1}$ and $X = X_{1}$, we obtain a state-space realization 
$\hat{\mathcal{B}} = \mathcal{B}_{s}^{(\mathbf{u},\mathbf{y})}$ with the required properties. \qed
\end{pf}

{\bf PROOF OF THEOREM \ref{thm:prprop}} (see p.\ \pageref{thm:prprop}). 
That \ref{nl:prbtc2} $\Rightarrow$ \ref{nl:prbtc1} follows from Theorem \ref{thm:ssssr} and \citep[Theorem 13]{THTPLSNA}, noting that condition 3 of that theorem holds with $X = I$. To see that \ref{nl:prbtc1} $\Rightarrow$ \ref{nl:prbtc2}, consider the realization in Lemma \ref{thm:ppst}. From that theorem, $(\hat{P}, \hat{Q})$ has a realization $(\tilde{A}, \tilde{B}, \tilde{C}, \tilde{D})$ with the following properties:
\begin{enumerate}[label=\arabic*., ref=\arabic*, leftmargin=0.5cm]
\item there exists $\tilde{X} \in \mathbb{R}^{d \times d}$ such that $\tilde{X} > 0$ and \newline
$\begin{bmatrix} -\tilde{X}\tilde{A} - \tilde{A}^{T}\tilde{X}& \tilde{C}^{T} - \tilde{X}\tilde{B}\\ \tilde{C} - \tilde{B}^{T}\tilde{X}& \tilde{D} + \tilde{D}^{T}\end{bmatrix} \geq 0$; and\label{nl:pprbtc1}
\item there exists a symmetric nonsingular $\tilde{S} \in \mathbb{R}^{d \times d}$ such that $\tilde{S}\tilde{A} = \tilde{A}^{T}\tilde{S}$, $\tilde{S}\tilde{B} = \tilde{C}^{T}$, and $\tilde{D} = \tilde{D}^{T}$.\label{nl:pprbtc2}
\end{enumerate}

Since $\tilde{X} > 0$, then there exists a nonsingular $\tilde{R} \in \mathbb{R}^{d \times d}$ such that $\tilde{X} = \tilde{R}^{T}\tilde{R}$. As $\tilde{S}$ is symmetric and nonsingular, then so too is $(\tilde{R}^{-1})^{T}\tilde{S}\tilde{R}^{-1}$. By considering an eigenvalue decomposition, we conclude that there exists a signature matrix $\Sigma_{i} = \text{diag}(I \hspace{0.25cm} {-}I) \in \mathbb{R}^{d \times d}$, a diagonal matrix $0 < W \in \mathbb{R}^{d \times d}$, and an orthogonal matrix $V \in \mathbb{R}^{d \times d}$ (i.e., $V^{T} = V^{-1}$), such that $(\tilde{R}^{-1})^{T}\tilde{S}\tilde{R}^{-1} = V\Sigma_{i} WV^{T}$. Here, $\Sigma_{i} W = W \Sigma_{i}$ is a diagonal matrix containing the eigenvalues of $(\tilde{R}^{-1})^{T}\tilde{S}\tilde{R}^{-1}$, which are necessarily real since $(\tilde{R}^{-1})^{T}\tilde{S}\tilde{R}^{-1}$ is symmetric. Now, let
\begin{equation*}
\hat{Y} = \left[\!\begin{smallmatrix}-\hat{A}& -\hat{B}\\ \hat{C}& \hat{D}\end{smallmatrix}\!\right] := \left[\!\begin{smallmatrix}V^{T}\tilde{R}& 0\\0& I\end{smallmatrix}\!\right]\left[\!\begin{smallmatrix}-\tilde{A}& -\tilde{B}\\ \tilde{C}& \tilde{D}\end{smallmatrix}\!\right]\left[\!\begin{smallmatrix}V^{T}\tilde{R}& 0\\0& I\end{smallmatrix}\!\right]^{-1}.
\end{equation*}
Then $(\hat{A}, \hat{B}, \hat{C}, \hat{D})$ is a realization for $(\hat{P}, \hat{Q})$, and
\begin{equation*}
\hat{Y} 
= \left[\!\begin{smallmatrix}\tilde{R}^{-1}V& 0\\0& I\end{smallmatrix}\!\right]^{T}\left[\!\begin{smallmatrix}-\tilde{X}\tilde{A}& -\tilde{X}\tilde{B}\\ \tilde{C}& \tilde{D}\end{smallmatrix}\!\right]\left[\!\begin{smallmatrix}\tilde{R}^{-1}V& 0\\0& I\end{smallmatrix}\!\right],
\end{equation*}
which implies that $\hat{Y} + \hat{Y}^{T} \geq 0$. Next, let
\begin{equation*}
Y = \left[\!\begin{smallmatrix}-A& -B\\ C& D\end{smallmatrix}\!\right] := \left[\!\begin{smallmatrix}W^{1/2}& 0\\0& I\end{smallmatrix}\!\right]\left[\!\begin{smallmatrix}-\hat{A}& -\hat{B}\\ \hat{C}& \hat{D}\end{smallmatrix}\!\right]\left[\!\begin{smallmatrix}W^{1/2}& 0\\0& I\end{smallmatrix}\!\right]^{-1}.
\end{equation*}
Then $(A, B, C, D)$ is also a realization for $(\hat{P}, \hat{Q})$. Also, with the notation $G := \tilde{R}^{-1}VW^{-1/2}$, then $G^{T}\tilde{S}G = W^{-1/2}V^{T}(\tilde{R}^{-1})^{T}\tilde{S}\tilde{R}^{-1}VW^{-1/2} = W^{-1/2}\Sigma_{i}WW^{-1/2} = \Sigma_{i}$ $A = G^{-1}\tilde{A}G, B = G^{-1}\tilde{B}$, and $C = \tilde{C}G$. Thus, $\Sigma_{i}A = G^{T}\tilde{S}\tilde{A}G = G^{T}\tilde{A}^{T}\tilde{S}G = A^{T}\Sigma_{i}$, $\Sigma_{i} B = G^{T}\tilde{S}\tilde{B} = G^{T}\tilde{C}^{T} = C^{T}$, and $D = D^{T}$, which implies that $\text{diag}({-}\Sigma_{i}\hspace{0.15cm} I)Y$ is symmetric. Note that $W^{1/2}$ is diagonal since $W$ is, and partition $W^{1/2}$ compatibly with $\Sigma_{i}$ as $W^{1/2} = \text{diag}(F_{1} \hspace{0.15cm} F_{2})$. Also, partition $Y$ and $\hat{Y}$ compatibly with $\text{diag}({-}\Sigma_{i} \hspace{0.15cm} I) = \text{diag}({-}I \hspace{0.15cm} I \hspace{0.15cm} I)$ as follows:
\begin{equation*}
Y = \left[\!\begin{smallmatrix}F_{1}& 0& 0 \\0& F_{2}& 0\\ 0& 0& I\end{smallmatrix}\!\right]\left[\!\begin{smallmatrix}-\hat{A}_{11}& -\hat{A}_{12}& -\hat{B}_{1}\\ -\hat{A}_{21}& -\hat{A}_{22}& -\hat{B}_{2}\\ \hat{C}_{1}& \hat{C}_{2}& \hat{D}\end{smallmatrix}\!\right]\left[\!\begin{smallmatrix}F_{1}^{-1}& 0& 0 \\0& F_{2}^{-1}& 0\\ 0& 0& I\end{smallmatrix}\!\right].
\end{equation*}
Then, let\begin{align*}
Z_{11} &{:=} -F_{1}\hat{A}_{11}F_{1}^{-1}, \hspace{0.15cm} Z_{22} {:=} \left[\!\begin{smallmatrix} F_{2}& 0\\ 0& I\end{smallmatrix}\!\right]\left[\!\begin{smallmatrix}-\hat{A}_{22}& -\hat{B}_{2}\\ \hat{C}_{2}& \hat{D}\end{smallmatrix}\!\right]\left[\!\begin{smallmatrix}F_{2}^{-1}& 0\\ 0& I\end{smallmatrix}\!\right] \\
Z_{12} &{:=} -F_{1}\left[\!\begin{smallmatrix}\hat{A}_{12}& \hat{B}_{1}\end{smallmatrix}\!\right]\left[\!\begin{smallmatrix}F_{2}^{-1}& 0\\ 0& I\end{smallmatrix}\!\right], \text{ and } Z_{21} {:=} \left[\!\begin{smallmatrix} F_{2}& 0\\ 0& I\end{smallmatrix}\!\right]\left[\!\begin{smallmatrix}-\hat{A}_{21}\\ \hat{C}_{1}\end{smallmatrix}\!\right]F_{1}^{-1}.
\end{align*}
Since $\text{diag}(-\Sigma_{i}\hspace{0.15cm} I)Y = \text{diag}({-}I \hspace{0.15cm} I \hspace{0.15cm} I)Y$ is symmetric, then we conclude that $Z_{11}$ and $Z_{22}$ are both symmetric, and $Z_{12} = -Z_{21}^{T}$. Thus, $Y+Y^{T} = \text{diag}(2Z_{11} \hspace{0.15cm} 2Z_{22})$, and to complete the proof it remains to show that $Z_{11} \geq 0$ and $Z_{22} \geq 0$. To prove this, we recall that $\hat{Y} + \hat{Y}^{T} \geq 0$, so
\begin{equation*}
-\hat{A}_{11} - \hat{A}_{11}^{T} \geq 0, \text{ and } \begin{bmatrix}-\hat{A}_{22}& -\hat{B}_{2}\\ \hat{C}_{2}& \hat{D}\end{bmatrix} + \begin{bmatrix}-\hat{A}_{22}& -\hat{B}_{2}\\ \hat{C}_{2}& \hat{D}\end{bmatrix}^{T} \geq 0.
\end{equation*}
Since $Z_{11}$ and $Z_{22}$ are both symmetric, then their eigenvalues are all real. Now, let $\lambda < 0$, and let $\mathbf{z}$ be a real vector with $Z_{11}\mathbf{z} = \lambda \mathbf{z}$. Then $\hat{\mathbf{z}} := F_{1}^{-1}\mathbf{z}$ satisfies $-\hat{A}_{11}\hat{\mathbf{z}} = \lambda \hat{\mathbf{z}}$. Thus, $\hat{\mathbf{z}}^{T}(-\hat{A}_{11} -\hat{A}_{11}^{T})\hat{\mathbf{z}} = 2\lambda \hat{\mathbf{z}}^{T}\hat{\mathbf{z}} \leq 0$. Since $(-\hat{A}_{11} -\hat{A}_{11}^{T}) \geq 0$, then we conclude that $\hat{\mathbf{z}} = 0$. It follows that the eigenvalues of $Z_{11}$ are all real and non-negative, whence $Z_{11} \geq 0$. A similar argument then shows that $Z_{22} \geq 0$, and completes the proof. \qed


{\bf PROOF OF THEOREM \ref{thm:prbtp1}} (see p.\ \pageref{thm:prbtp1}). That \ref{nl:prbtp1c1} $\iff$ \ref{nl:prbtp1c2} follows from \cite[Theorem 9]{THTPLSNA} and Theorem \ref{thm:sigsymrep} of the present paper.

If $\mathcal{B}$ takes the form of $\hat{\mathcal{B}}$ in (\ref{eq:bgsqd}), then from Theorem \ref{thm:prprop} it follows that $\mathcal{B}$ is passive and reciprocal if and only if $\mathcal{B}$ has a state-space realization with the properties outlined in condition \ref{nl:prbtc2} of that theorem. From \citep[Sections 4.4 and 9.4]{AndVong}, this holds if and only if $\mathcal{B}$ is the driving-point behavior of an RLCT network. It remains to consider the case in which $\mathcal{B}$ does not take the form of $\hat{\mathcal{B}}$ in (\ref{eq:bgsqd}), i.e., $P, Q$ in (\ref{eq:bgd}) are such that $Q$ is singular or $Q^{-1}P$ is not proper.

{\bf  \ref{nl:prbtp1c3} $\boldsymbol{\Rightarrow}$ \ref{nl:prbtp1c1}.} \hspace{0.3cm} That $\mathcal{B}$ is passive follows from \citep[Theorem 6]{HUGIFAC}. It remains to show that $\mathcal{B}$ is reciprocal. As explained in \citep[Section 2]{HUGIFAC}, any given RLCT network corresponds to a cascade loading of two networks: (i) $N_{1}$, in which all of the elements (resistors, inductors, capacitors and transformers) are removed and every single element port is replaced with an external port; and (ii) $N_{2}$, which contains each of the elements in the original circuit (disconnected from each other). Furthermore, the driving-point behaviors of $N_{1}$ and $N_{2}$ are both reciprocal.\footnote{To see this, note initially that the behavior of the network $N_{1}$ has the same form as the driving-point behavior of a transformer \citep{AndNew}. It is then easily verified that the driving-point behaviors of resistors, inductors, capacitors and transformers are all reciprocal, and so too are the driving-point behaviors of $N_{1}$ and $N_{2}$.}  
Now, let $\mathcal{B}$ and $\tilde{\mathcal{B}}$ be fixed but arbitrary reciprocal behaviors, and let (i) $(\text{col}(\mathbf{i}_{a,1} \hspace{0.15cm} \mathbf{i}_{a,2}), \text{col}(\mathbf{v}_{a,1} \hspace{0.15cm} \mathbf{v}_{a,2})) \in \mathcal{B}$, $(\mathbf{i}_{a,3}, \mathbf{v}_{a,3}) \in \tilde{\mathcal{B}}$, $(\text{col}(\mathbf{i}_{b,1} \hspace{0.15cm} \mathbf{i}_{b,2}), \text{col}(\mathbf{v}_{b,1} \hspace{0.15cm} \mathbf{v}_{b,2})) \in \mathcal{B}$, and $(\mathbf{i}_{b,3}, \mathbf{v}_{b,3}) \in \tilde{\mathcal{B}}$; (ii) $\mathbf{i}_{a,3} = -\mathbf{i}_{a,2}$, $\mathbf{v}_{a,3} = \mathbf{v}_{a,2}$, $\mathbf{i}_{b,3} = -\mathbf{i}_{b,2}$, and $\mathbf{v}_{b,3} = \mathbf{v}_{b,2}$; and (iii) $\mathbf{i}_{a,1}, \mathbf{i}_{a,2}, \mathbf{v}_{a,1}, \mathbf{v}_{a,2}, \mathbf{i}_{b,1}, \mathbf{i}_{b,2}, \mathbf{v}_{b,1}$, and $\mathbf{v}_{b,2}$ have compact support on the left. 
Then it suffices to show that $\mathbf{v}_{b,1} \star \mathbf{i}_{a,1} = \mathbf{i}_{b,1} \star \mathbf{v}_{a,1}$. To prove this, note that, since $\mathcal{B}$ and $\tilde{\mathcal{B}}$ are reciprocal, then 
\begin{align*}
\mathbf{v}_{b,1} \star \mathbf{i}_{a,1} + \mathbf{v}_{b,2} \star \mathbf{i}_{a,2} - \mathbf{i}_{b,1} \star \mathbf{v}_{a,1}  - \mathbf{i}_{b,2} \star \mathbf{v}_{a,2} = 0, \\
\text{and } \mathbf{v}_{b,3} \star \mathbf{i}_{a,3} - \mathbf{i}_{b,3} \star \mathbf{v}_{a,3} = 0,
\end{align*}
whence $\mathbf{v}_{b,1} \star \mathbf{i}_{a,1} - \mathbf{i}_{b,1} \star \mathbf{v}_{a,1} = 0$.


{\bf \ref{nl:prbtp1c2} $\boldsymbol{\Rightarrow}$ \ref{nl:prbtp1c3}.} \hspace{0.3cm} We will show the following:
\begin{enumerate}[label=(\alph*), ref=(\alph*), leftmargin=0.5cm]
\item If $\lambda_{0} \in \mathbb{C}_{+}$ and $\mathbf{z} \in \mathbb{C}^{n}$ satisfy $Q(\lambda_{0})\mathbf{z} = 0$, then $Q\mathbf{z} = 0$.\label{prbtp1tc1} 
\item There exists a nonsingular matrix $T \in \mathbb{R}^{n \times n}$ and a unimodular matrix $\hat{Y}$ such that
\begin{equation*}
\begin{bmatrix}P& -Q\end{bmatrix} = \hat{Y}\begin{bmatrix}\hat{P}& - \hat{Q}\end{bmatrix}\begin{bmatrix}T& 0\\0& (T^{T})^{-1}\end{bmatrix},
\end{equation*}
where $\hat{P}$ and $\hat{Q}$ have the compatible partitions
\begin{equation*}
\hat{P}  = \begin{bmatrix}\hat{P}_{11}& 0\\0& I\end{bmatrix} \text{ and } \hat{Q} = \begin{bmatrix}\hat{Q}_{11}& 0\\ 0& 0\end{bmatrix},
\end{equation*}
and where $\hat{Q}_{11}$ is nonsingular, $\hat{Q}_{11}^{-1}\hat{P}_{11}$ is symmetric, and $(\hat{P}_{11}, \hat{Q}_{11})$ is a positive-real pair.\label{prbtp1tc2}
\item With $\hat{P}_{11}$ and $\hat{Q}_{11}$ as in \ref{prbtp1tc2}, then the limit $\lim_{\xi \rightarrow \infty}((1/\xi)(\hat{Q}_{11}^{-1}\hat{P}_{11})(\xi))$ exists and is non-negative definite. Also, with the notation $K := \lim_{\xi \rightarrow \infty}((1/\xi)(\hat{Q}_{11}^{-1}\hat{P}_{11})(\xi))$, $\tilde{P}(\xi) := \hat{P}_{11}(\xi) - \hat{Q}_{11}(\xi)K\xi$, and $\tilde{Q}(\xi) := \hat{Q}_{11}(\xi)$, then $\tilde{Q}^{-1}\tilde{P}$ is proper and symmetric and $(\tilde{P}, \tilde{Q})$ is a positive-real pair.\label{prbtp1tc3}
\end{enumerate}

\begin{figure}[!t]
\scriptsize
 \begin{center}
\leavevmode
\includegraphics[width=1.0\hsize]{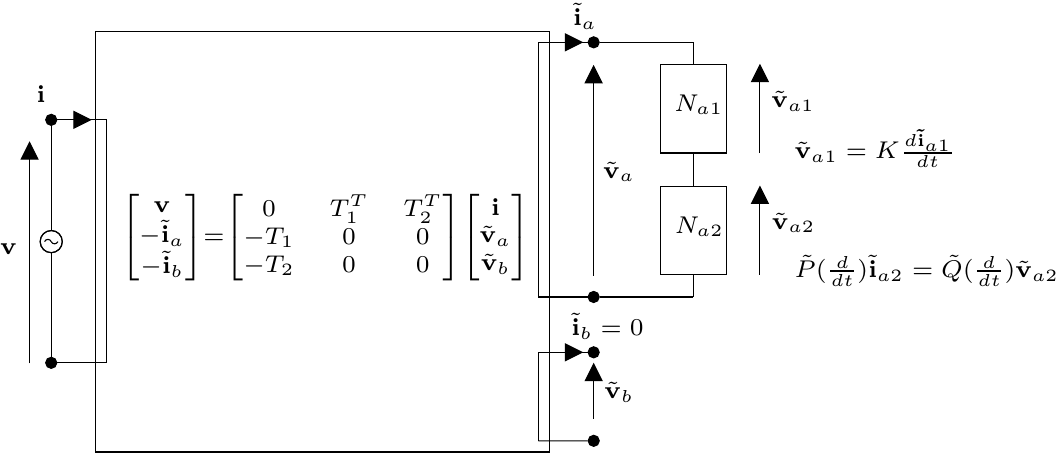}
\end{center}
\caption{RLCT network realization of the behavior in (\ref{eq:rlct_gen}).}
\label{fig:rlct_gen}
\end{figure}

Now, let $\tilde{P}, \tilde{Q}$ and $K$ be as defined in \ref{prbtp1tc3}. It follows from Theorem \ref{thm:prprop} and \citep[Sections 4.4 and 9.4]{AndVong} that there exist RLCT networks $N_{1}$ and $N_{2}$ whose driving-point behaviors take the form $\lbrace (\mathbf{i}, \mathbf{v}) \in \mathcal{L}_{1}^{\text{loc}}\left(\mathbb{R}, \mathbb{R}^{\tilde{n}}\right) {\times} \mathcal{L}_{1}^{\text{loc}}\left(\mathbb{R}, \mathbb{R}^{\tilde{n}}\right) \mid \tilde{P}(\tfrac{d}{dt})\mathbf{i} = \tilde{Q}(\tfrac{d}{dt})\mathbf{v}\rbrace$ and $\lbrace (\mathbf{i}, \mathbf{v}) \in \mathcal{L}_{1}^{\text{loc}}\left(\mathbb{R}, \mathbb{R}^{\tilde{n}}\right) {\times} \mathcal{L}_{1}^{\text{loc}}\left(\mathbb{R}, \mathbb{R}^{\tilde{n}}\right) \mid K\tfrac{d\mathbf{i}}{dt} = \mathbf{v}\rbrace$, respectively. Next, let $T$ be as in \ref{prbtp1tc2}, partition $T$ and $T^{-1}$ compatibly with $\hat{P}$ as $T =  \text{col}(T_{1} \hspace{0.15cm} T_{2})$ and $T^{-1} = [\hat{T}_{1} \hspace{0.15cm} \hat{T}_{2}]$, and consider the behavior corresponding to the set of locally integrable solutions to
\begin{equation}
\left[\!\begin{smallmatrix}0& 0& 0& 0& \tilde{P}(\tfrac{d}{dt})& 0& 0& {-}\tilde{Q}(\tfrac{d}{dt})\\
0& 0& 0& 0& 0& I& 0& 0\\
0& I& {-}T_{1}^{T}& {-}T_{2}^{T}& 0& 0& 0& 0\\
T_{1}& 0& 0& 0& {-}I& 0& 0& 0\\
T_{2}& 0& 0& 0& 0& {-}I& 0& 0\\
0& 0& 0& 0& K\tfrac{d}{dt}& 0& {-}I& 0\\
0& 0& I& 0& 0& 0& {-}I& {-}I\end{smallmatrix}\right]\! \!\left[\!\begin{smallmatrix}\mathbf{i}\\ \mathbf{v}\\ \tilde{\mathbf{v}}_{a}\\ \tilde{\mathbf{v}}_{b}\\ \tilde{\mathbf{i}}_{a}\\ \tilde{\mathbf{i}}_{b}\\ \tilde{\mathbf{v}}_{a1}\\ \tilde{\mathbf{v}}_{a2}\end{smallmatrix}\!\right] {=} 0,\label{eq:rlct_gen}
\end{equation}
which is the driving-point behavior of the RLCT network in Fig.\ \ref{fig:rlct_gen}. We then let
\begin{align*}
U &:= \begin{bmatrix}\hat{Y}& 0\\ 0& I\end{bmatrix}\begin{bmatrix}I& Z\\ 0& I\end{bmatrix}, \text{ with} \\
Z(\xi) &:= \begin{bmatrix}{-}\tilde{Q}(\xi)\hat{T}_{1}^{T}& \tilde{P}(\xi) {+} \tilde{Q}(\xi)K\xi& 0& \tilde{Q}(\xi)& {-}\tilde{Q}(\xi)\\
0& 0& I& 0& 0\end{bmatrix},
\end{align*}
and it is clear that $U$ is unimodular. Then, following Appendix \ref{app:b}, we pre-multiply both sides in (\ref{eq:rlct_gen}) by $U(\tfrac{d}{dt})$, we note that $\hat{T}_{1}^{T}T_{1}^{T} = I$ and $\hat{T}_{1}^{T}T_{2}^{T} = 0$, and we find that the driving-point behavior of $N$ is the set of locally integrable solutions to the differential equation
\begin{equation*}
\hat{Y}\begin{bmatrix}(\tilde{P}(\tfrac{d}{dt}) + \tilde{Q}(\tfrac{d}{dt})K\tfrac{d}{dt})T_{1}& -\tilde{Q}(\tfrac{d}{dt})\hat{T}_{1}^{T}\\ T_{2}& 0\end{bmatrix}\begin{bmatrix}\mathbf{i}\\ \mathbf{v}\end{bmatrix} = 0.
\end{equation*}
But from \ref{prbtp1tc2} and \ref{prbtp1tc3}, the leftmost matrix in this equation is equal to $[P \hspace{0.25cm} {-}Q](\tfrac{d}{dt})$. In other words, the driving-point behavior of $N$ is $\mathcal{B}$.

It remains to show conditions \ref{prbtp1tc1}--\ref{prbtp1tc3}. To show condition \ref{prbtp1tc1}, we let $\hat{P}:=P-Q$ and $\hat{Q} := P+Q$. Since $(P,Q)$ is a positive-real pair, then $\hat{Q}(\lambda)\hat{Q}(\bar{\lambda})^{T} - \hat{P}(\lambda)\hat{P}(\bar{\lambda})^{T} = 2(P(\lambda)Q(\bar{\lambda})^{T} + Q(\lambda)P(\bar{\lambda})^{T}) \geq 0$ for all $\lambda \in \mathbb{C}_{+}$. Now, suppose $\lambda_{0} \in \mathbb{C}_{+}$ and $\mathbf{w} \in \mathbb{C}^{n}$ satisfy $\mathbf{w}^{T}\hat{Q}(\lambda_{0}) = 0$. Then $-\mathbf{w}^{T}\hat{P}(\lambda_{0})\hat{P}(\bar{\lambda}_{0})^{T}\bar{\mathbf{w}} \geq 0$, which implies that $\mathbf{w}^{T}\hat{P}(\lambda_{0}) = 0$. But this implies that $\mathbf{w}^{T}[P \hspace{0.25cm} {-}Q](\lambda_{0}) = 0$, whence $\mathbf{w} = 0$ since $(P,Q)$ is a positive-real pair. We conclude that $\hat{Q}(\lambda)$ is nonsingular for all $\lambda \in \mathbb{C}_{+}$, and so $I - (\hat{Q}^{-1}\hat{P})(\lambda)((\hat{Q}^{-1}\hat{P})(\bar{\lambda}))^{T} \geq 0$ for all $\lambda \in \mathbb{C}_{+}$. This implies that $(\hat{Q}^{-1}\hat{P})^{T}$ is bounded-real in accordance with \citep[Definition 16]{Youlaflpns}, and so $\hat{Q}^{-1}\hat{P}$ is bounded-real by \citep[Corollary 7(c)]{Youlaflpns}. It then follows from \citep[proof of Theorem 7]{Youlaflpns} that, if $\lambda_{0} \in \mathbb{C}_{+}$ and $\mathbf{w} \in \mathbb{C}^{n}$ satisfy $(I - (\hat{Q}^{-1}\hat{P})(\lambda_{0}))\mathbf{w} = 0$, then $(I - \hat{Q}^{-1}\hat{P})\mathbf{w} = 0$. Now, let $\lambda_{0} \in \mathbb{C}_{+}$ and $\mathbf{z} \in \mathbb{C}^{n}$ satisfy $Q(\lambda_{0})\mathbf{z} = 0$. Then $(P+Q)^{-1}(\lambda_{0})Q(\lambda_{0})\mathbf{z} = \tfrac{1}{2}(I - (P+Q)^{-1}(\lambda_{0})(P-Q)(\lambda_{0}))\mathbf{z} = 0$, whence $(I - (P+Q)^{-1}(P-Q))\mathbf{z} = 0$, and so $Q\mathbf{z} = \tfrac{1}{2}(P+Q)(I - (P+Q)^{-1}(P-Q))\mathbf{z} = 0$.

To show condition \ref{prbtp1tc2}, we let $r := \text{normalrank}(Q)$, and we first show that there exists a nonsingular matrix $T = \text{col}(T_{1}\hspace{0.15cm} T_{2}) \in \mathbb{R}^{n \times n}$ with $T_{1} \in \mathbb{R}^{r \times n}$ such that $QT_{2}^{T} = 0$. Accordingly, we let the columns of $W \in \mathbb{R}^{n \times (n{-}r)}[\xi]$ be a basis for the right syzygy of $Q$ \citep[see][p.\ 85]{JWBAOIS}, and we pick a fixed but arbitrary $\lambda_{0} > 0$. Then $W(\lambda_{0}) \in \mathbb{R}^{n \times (n{-}r)}$ has full column rank and $Q(\lambda_{0})W(\lambda_{0}) = 0$, whence $QW(\lambda_{0}) = 0$ by condition \ref{prbtp1tc1}. We then let $T$ be a nonsingular matrix whose final $n{-}r$ rows are $W(\lambda_{0})^{T}$.

Next, note that $QT_{1}^{T} \in \mathbb{R}^{n \times r}$ and $\text{normalrank}(QT_{1}^{T}) = \text{normalrank}(QT^{T}) = r$. Then, by considering the upper echelon form for $QT_{1}^{T}$ \citep[see][Note A4]{THTPLSNA}, we obtain a unimodular $Y \in \mathbb{R}^{n \times n}$ such that $YQT^{T} =: \hat{Q}$ takes the form indicated in condition \ref{prbtp1tc2}, where $\hat{Q}_{11} \in \mathbb{R}^{r \times r}[\xi]$ is nonsingular. Now, let $\hat{P} := YPT^{-1}$. It is then easily shown that $(\hat{P}, \hat{Q})$ is a positive-real pair since $(P,Q)$ is. Accordingly, we consider a fixed but arbitrary $\lambda \in \mathbb{C}_{+}$, we partition $\hat{P}$ compatibly with $\hat{Q}$ as
\begin{equation*}
\hat{P} = \begin{bmatrix}\hat{P}_{11}& \hat{P}_{12}\\ \hat{P}_{21}& \hat{P}_{22}\end{bmatrix},
\end{equation*}
and it follows that
\begin{equation*}
\hspace*{-0.3cm}\begin{bmatrix}\!\hat{P}_{11}(\lambda)(\hat{Q}_{11}(\bar{\lambda}))^{T} {+} \hat{Q}_{11}(\lambda)(\hat{P}_{11}(\bar{\lambda}))^{T}& \hat{Q}_{11}(\lambda)(\hat{P}_{21}(\bar{\lambda}))^{T}\\ \hat{P}_{21}(\lambda)(\hat{Q}_{11}(\bar{\lambda}))^{T}& 0\!\end{bmatrix} {\geq} 0.
\end{equation*}
This implies that $\hat{P}_{21}(\lambda)(\hat{Q}_{11}(\bar{\lambda}))^{T} = 0$. Since this relationship holds for all $\lambda \in \mathbb{C}_{+}$, and $\hat{Q}_{11}$ is nonsingular, then we conclude that $\hat{P}_{21} = 0$.

Next, it follows from \citep[proof of Lemma D.3 condition 1]{THAS} that $\hat{P}_{22}$ is unimodular since $(\hat{P}, \hat{Q})$ is a positive-real pair. Accordingly, we let
\begin{equation*}
\hat{Y} = Y^{-1}\begin{bmatrix}I& \hat{P}_{12}\\ 0& \hat{P}_{22}\end{bmatrix},
\end{equation*}
and we find that $\hat{Y}$ is unimodular and $[P \hspace{0.25cm} {-}Q]$ has the form indicated in condition \ref{prbtp1tc2}. Finally, it is easily shown that $(\hat{P}_{11}, \hat{Q}_{11})$ is a positive-real pair, and $\hat{P}\hat{Q}^{T} - \hat{Q}\hat{P}^{T} = \hat{Y}^{-1}(PQ^{T} - QP^{T})(\hat{Y}^{-1})^{T} = 0$ so $\hat{Q}_{11}^{-1}\hat{P}_{11}$ is symmetric.

The proof of condition \ref{prbtp1tc3} follows from \cite[Proof of Lemma D.4]{THAS}, noting in addition that $(\tilde{Q}^{-1}\tilde{P})(\xi) = (\hat{Q}_{11}^{-1}\hat{P}_{11})(\xi) - K\xi$, which is symmetric since $\hat{Q}_{11}^{-1}\hat{P}_{11}$ and $K$ are symmetric. \qed


\section{Examples}
\label{sec:ex}
\begin{figure}[!t]
\scriptsize
 \begin{center}
\leavevmode
\includegraphics[width=1.0\hsize]{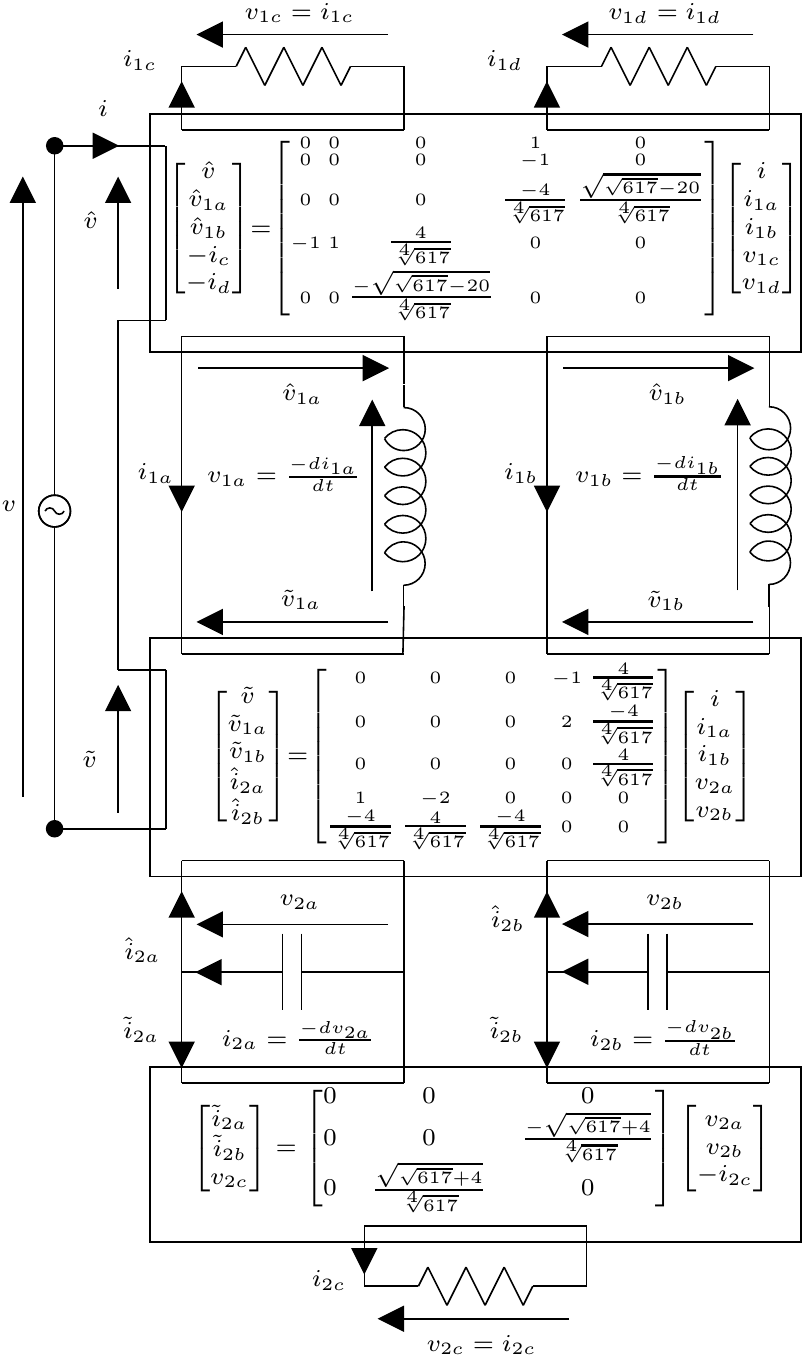}
\end{center}
\caption{RLCT network realization example 1.}
\label{fig:ex1}
\end{figure}
\begin{figure}[!t]
\scriptsize
 \begin{center}
\leavevmode
\includegraphics[width=1.0\hsize]{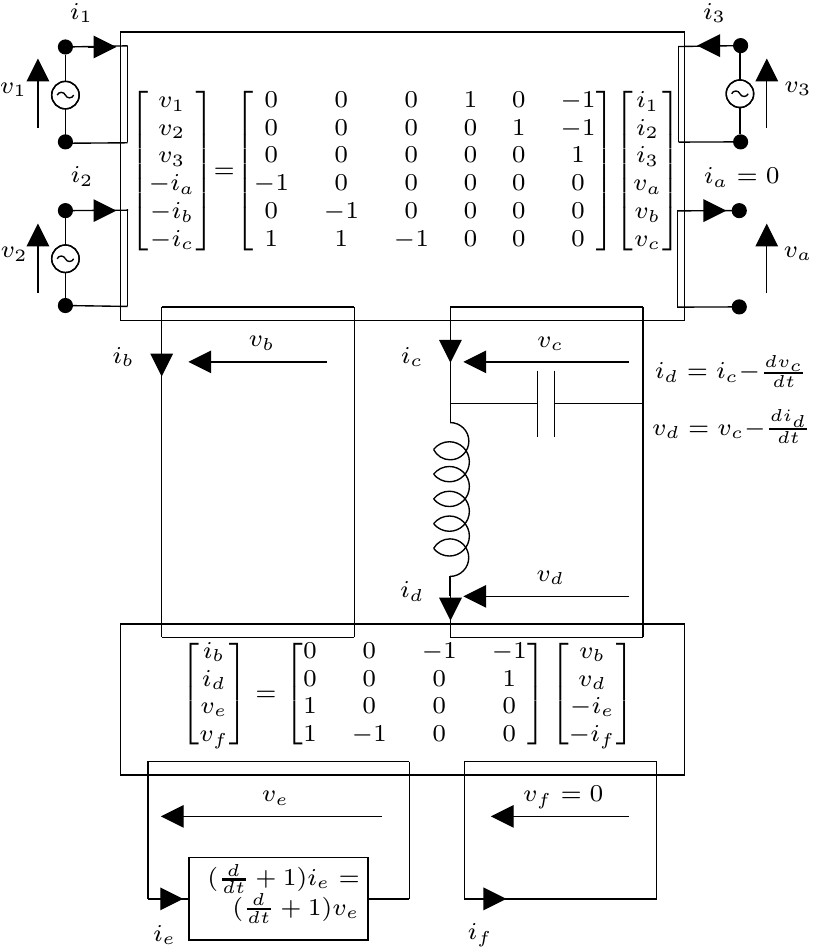}
\end{center}
\caption{RLCT network realization example 2.}
\label{fig:ex2}
\end{figure}
First, consider the problem of realising the driving-point behavior of the Bott Duffin networks in Fig.\ \ref{fig:bdne}. From Lemma \ref{lem:ssr}, we find that this corresponds to the set of solutions to the differential equation:
\begin{equation}
(\tfrac{d}{dt}+1)(\tfrac{d^2}{dt^2}+\tfrac{d}{dt}+1)i = (\tfrac{d}{dt}+1)(\tfrac{d^2}{dt^2}+\tfrac{d}{dt}+4)v.\label{eq:bdedpb}
\end{equation}
We begin by finding a $\mathcal{B}_{s}$ as in (\ref{eq:bhssr}) and matrices $X, S \in \mathbb{R}^{d \times d}$ as in Lemma \ref{thm:ppst}, and we then find a passive and reciprocal realization for this behavior as in Theorem \ref{thm:prprop}. Finally, we obtain an RLCT network realization from this passive and reciprocal realization using results in \cite{AndVong}. This realization procedure works in the general case, and relies on algorithms for: 1.\ computing a state-space realization for a behavior (see Remark \ref{rem:ssr}); 2.\ computing the available energy of a passive system (see \citep[Remark 15]{THAS}); 3.\ computing the null space and column space of a real matrix; 4.\ computing a Cholesky decomposition of a positive-definite matrix; 5.\ computing an eigenvalue decomposition of a symmetric matrix; and 6.\ solving Lyapunov and Sylvester equations \citep[as in][Theorems 3.7.3 and 3.7.4]{AndVong}.

We first obtain a state-space realization for the behavior in (\ref{eq:bdedpb}), and we transform this into controller staircase form \citep[see][Corollary 5.2.25]{JWIMTSC}. In this case, we find that $\mathcal{B}$ has a state-space realization $\mathcal{B} = \mathcal{B}_{s}^{(i,v)}$, where $\mathcal{B}_{s}$ is the set of solutions to
\begin{align*}
\tfrac{d\mathbf{x}}{dt} &= \tilde{A}\mathbf{x} + \tilde{B}i, \hspace{0.1cm} v = \tilde{C}\mathbf{x} + Di, \text{ where}\\
\tilde{A} &= \left[\!\begin{smallmatrix}0& 1& 0\\ -4& -1& 0\\ 0& 0& -1\end{smallmatrix}\!\right], \hspace{0.05cm} \tilde{B} = \left[\!\begin{smallmatrix}0\\ 1\\ 0\end{smallmatrix}\!\right], \hspace{0.05cm} \tilde{C} = \left[\!\begin{smallmatrix}-3& 0& 1\end{smallmatrix}\!\right], \hspace{0.05cm} D = 1.
\end{align*} 
Then, following Lemma \ref{thm:obspds}, we let $\Upsilon(K)$ and $A_{\Upsilon}(K)$ be as in (\ref{eq:upapsid}) with $\tilde{A}, \tilde{B}$ and $\tilde{C}$ substituted for $A, B$ and $C$, and we obtain symmetric matrices
\begin{equation*}
K_{1} = \frac{1}{4}\left[\!\begin{smallmatrix}1& -1& - 1\\ -1& 5& 5\\ -1& 5& 9\end{smallmatrix}\!\right], \text{ and } K_{-} = \frac{1}{9}\left[\!\begin{smallmatrix}2& -1& 0\\ - 1& 5& 0\\ 0& 0& 0\end{smallmatrix}\!\right],
\end{equation*}
where $K_{1} > 0$, $K_{-} \geq 0$, $\Upsilon(K_{1}) \geq 0$, $\Upsilon(K_{-}) \geq 0$, and $\text{spec}(A_{\Upsilon}(K_{-})) \in \bar{\mathbb{C}}_{-}$. Here, $K_{1}$ is obtained by computing the available energy for the system $\tfrac{d\mathbf{x}}{dt} = A\mathbf{x} + Bi, v = C\mathbf{x} + Di$ to obtain the matrix $X$ in Lemma \ref{thm:obspds}. Also, $K_{-}$ can be obtained by computing the available energy for the system $\tfrac{d\mathbf{\hat{x}}}{dt} = A^{T}\mathbf{\hat{x}} - C^{T}u, y = -B^{T}\mathbf{\hat{x}} + D^{T}u$ (see Lemma \ref{thm:obspds}). Next, note from Lemma \ref{thm:obspds} that there exists $\alpha > 0$ such that, for any given $0 < \epsilon \leq \alpha$, then $K_{\epsilon} \coloneqq \epsilon K_{1} + (1- \epsilon) K_{-}$ satisfies $K_{\epsilon} > 0$, $\Upsilon(K_{\epsilon}) \geq 0$, and $\text{spec}(A_{\Upsilon}(K_{\epsilon})) \in \bar{\mathbb{C}}_{-}$ . In this case,
\begin{equation*}
K = \frac{3K_{-} + K_{1}}{4} = \frac{1}{48}\left[\!\begin{smallmatrix}11& -7& -3\\ -7& 35& 15\\ -3& 15& 27\end{smallmatrix}\!\right],
\end{equation*}
and it can be verified that $K > 0$, $\Upsilon(K) \geq 0$, and $\text{spec}(A_{\Upsilon}(K)) \in \bar{\mathbb{C}}_{-}$. Now, following the proof of Theorem \ref{thm:ssssr}, we augment the matrices $\tilde{A}, \tilde{B}$ and $\tilde{C}$ to obtain an unobservable state-space realization $\mathcal{B} = \hat{\mathcal{B}}_{s}^{(i,v)}$, where $\hat{\mathcal{B}}_{s}$ corresponds to the set of solutions to the equation
\begin{align*}
\tfrac{d\hat{\mathbf{x}}}{dt} &= \hat{A}\hat{\mathbf{x}} + \hat{B}i, \hspace{0.1cm} v = \hat{C}\hat{\mathbf{x}} + Di, \text{ where} \displaybreak[3]\\
\hat{A} &= \left[\!\begin{smallmatrix}0& 1& 0& 0\\ -4& -1& 0& 0\\ 0& 0& -1& 0\\ 0& 0& 0& -1\end{smallmatrix}\!\right], \hspace{0.05cm} \hat{B} = \left[\!\begin{smallmatrix}0\\ 1\\ 0\\ 1\end{smallmatrix}\!\right], \hspace{0.05cm} \hat{C} = \left[\!\begin{smallmatrix}-3& 0& 1& 0\end{smallmatrix}\!\right], 
\end{align*}
and, as before, $D = 1$. Then, in Lemma \ref{lem:pdrr}, we let $\tilde{A}_{11}, \tilde{A}_{21}, \tilde{A}_{22}, \tilde{B}_{1}, \tilde{B}_{2}$ and $\tilde{C}_{1}$ be obtained by partitioning $\hat{A}, \hat{B}$ and $\hat{C}$ (here, $\tilde{A}_{11}$ is formed from the first three rows and columns of $\hat{A}$, and so forth), and we find that $K_{12}^{T} = \tfrac{1}{18}[7 \hspace{0.15cm} -11 \hspace{0.15cm} -9]$ solves the Sylvester equation in that lemma, and $\nabla = \tfrac{1}{2}$ solves the Lyapunov equation $\Psi(\nabla) = 0$ in that lemma. Accordingly, we obtain
\begin{equation*}
\hat{K} = \left[\!\begin{smallmatrix}\tfrac{11}{48}& -\tfrac{7}{48}& -\tfrac{1}{16}& \tfrac{7}{18}\\ -\tfrac{7}{48}& \tfrac{35}{48}& \tfrac{5}{16}& -\tfrac{11}{18}\\ -\tfrac{1}{16}& \tfrac{5}{16}& \tfrac{9}{16}& -\tfrac{1}{2}\\ \tfrac{7}{18}& -\tfrac{11}{18}& -\tfrac{1}{2}& \tfrac{16}{9}\end{smallmatrix}\!\right],
\end{equation*}
which satisfies $\hat{K} > 0$ and $-\hat{K}\hat{A}^{T} - \hat{A}\hat{K} - (\hat{K}\hat{C}^{T} - \hat{B})(D+D^{T})^{-1}(\hat{C}\hat{K} - \hat{B}^{T})$. We then let $\hat{X} = \hat{K}^{-1}$. Also, following Remark \ref{rem:compsm}, we obtain the matrix
\begin{equation*}
\hat{S} = \left[\!\begin{smallmatrix} -3& -3& 0& 0\\ -3& 0& 0& 0\\ 0& 0& 0& 1\\ 0& 0& 1& 0\end{smallmatrix}\!\right],
\end{equation*}
which satisfies $\hat{S}\hat{A} = \hat{A}^{T}\hat{S}$ and $\hat{S}\hat{B} = \hat{C}^{T}$. We have thus obtained a state-space realization $\mathcal{B}_{s}$ as in (\ref{eq:bhssr}) and matrices $X, S \in \mathbb{R}^{d \times d}$ as in Lemma \ref{thm:ppst}.

Next, using a Cholesky decomposition we obtain
\begin{equation*}
\hat{X} = R^{T}R, \text{ with } R = \left[\!\begin{smallmatrix}\tfrac{\sqrt{357}}{7}& \tfrac{\sqrt{3}}{\sqrt{119}}& -\tfrac{20}{3\sqrt{357}}& -\tfrac{4\sqrt{3}}{\sqrt{119}}\\ 0& \tfrac{6}{\sqrt{17}}& -\tfrac{2}{\sqrt{17}}& \tfrac{3}{2\sqrt{17}}\\ 0& 0& \tfrac{8}{3\sqrt{3}}& \tfrac{\sqrt{3}}{4}\\ 0& 0& 0& \tfrac{3}{4}\end{smallmatrix}\!\right].
\end{equation*}
Then, following the proof of Theorem \ref{thm:prprop}, we compute an eigenvalue decomposition of $(R^{-1})^{T}\hat{S}R^{-1}$ to obtain
\begin{align*}
\Sigma_{i} &= \left[\!\begin{smallmatrix}-1 & 0& 0& 0\\ 0& -1& 0& 0\\ 0& 0& 1& 0\\ 0& 0& 0& 1\end{smallmatrix} \!\right], \hspace{0.05cm} W = \left[\!\begin{smallmatrix}1& 0 & 0& 0\\ 0& \tfrac{\sqrt{617}+13}{32}& 0& 0\\ 0& 0& 1& 0\\ 0& 0& 0& \tfrac{\sqrt{617}-13}{32}\end{smallmatrix}\!\right], \text{ and} \displaybreak[3] \\
V &= \left[\!\begin{smallmatrix}\tfrac{2\sqrt{7}}{\sqrt{51}}& \tfrac{-2\sqrt{2}\sqrt{\sqrt{617}-13}}{\sqrt{51}\sqrt[4]{617}}& \tfrac{\sqrt{7}}{\sqrt{51}}& \tfrac{-2\sqrt{2}\sqrt{\sqrt{617}+13}}{\sqrt{51}\sqrt[4]{617}}\\ \tfrac{5}{2\sqrt{17}}& \tfrac{\sqrt{7}\sqrt{\sqrt{617}-13}}{2\sqrt{34}\sqrt[4]{617}}& -\tfrac{3}{\sqrt{17}} & \tfrac{\sqrt{7}\sqrt{\sqrt{617}+13}}{2\sqrt{34}\sqrt[4]{617}}\\ \tfrac{1}{4\sqrt{3}}& -\tfrac{\sqrt{43\sqrt{617}-295}}{4\sqrt{6}\sqrt[4]{617}}& \tfrac{1}{2\sqrt{3}}&  \tfrac{\sqrt{43\sqrt{617}+295}}{4\sqrt{6}\sqrt[4]{617}}\\ \tfrac{1}{4}& \tfrac{\sqrt{11\sqrt{617}+185}}{4\sqrt{2}\sqrt[4]{617}}& \tfrac{1}{2}&  \tfrac{\sqrt{11\sqrt{617}-185}}{4\sqrt{2}\sqrt[4]{617}}\end{smallmatrix}\!\right].
\end{align*}
Following the proof of Theorem \ref{thm:prprop}, let $G := R^{-1}VW^{-1/2}$, $A = G^{-1}\hat{A}G, B = G^{-1}\hat{B}$, and $C = \hat{C}G$, which gives
\begin{align*}
A = \left[\!\begin{smallmatrix}-1& \tfrac{-4}{\sqrt[4]{617}}& -2& \tfrac{4}{\sqrt[4]{617}}\\ \tfrac{-4}{\sqrt[4]{617}}& \tfrac{4}{\sqrt{617}}-1& 0& \tfrac{-4}{\sqrt{617}}\\ 2& 0& 0& 0\\ \tfrac{-4}{\sqrt[4]{617}}& \tfrac{4}{\sqrt{617}}& 0& -1-\tfrac{4}{\sqrt{617}}\end{smallmatrix}\!\right], \hspace{0.05cm} B = \left[\!\begin{smallmatrix}1\\ \tfrac{4}{\sqrt[4]{617}}\\ -1\\ \tfrac{4}{\sqrt[4]{617}}\end{smallmatrix}\!\right], \text{ and}
\end{align*}
$C = B^{T}\Sigma_{i}$. 
These satisfy the conditions of Theorem \ref{thm:prprop}. 

Next, we use the results in \citep[Sections 9.2 and 9.4]{AndVong} to obtain an RLCT network which realizes the behavior in (\ref{eq:bdedpb}). We let
\begin{equation*}
M:=\begin{bmatrix}D& C\\-B& -A\end{bmatrix}, \text{ and } \Sigma = \begin{bmatrix}1& 0\\0& -\Sigma_{i}\end{bmatrix},
\end{equation*}
and we conclude that $M + M^{T} \geq 0$ and $\Sigma M$ is symmetric. It follows that $M$ takes the form
\begin{equation*}
M = \begin{bmatrix}M_{11}& -M_{21}^{T}\\ M_{21}& M_{22}\end{bmatrix},
\end{equation*}
where $M_{11} \in \mathbb{R}^{3 \times 3}$ and $M_{22} \in \mathbb{R}^{2 \times 2}$ are symmetric, and $M_{11}, M_{22} \geq 0$. In this case, we have
\begin{align*}
M_{11} &= 
\left[\!\begin{smallmatrix}1& 0\\ -1& 0\\ \tfrac{-4}{\sqrt[4]{617}} & \tfrac{\sqrt{\sqrt{617}-20}}{\sqrt[4]{617}}\end{smallmatrix}\!\right]\left[\!\begin{smallmatrix}1& -1& \tfrac{-4}{\sqrt[4]{617}}\\ 0&0 & \tfrac{\sqrt{\sqrt{617}-20}}{\sqrt[4]{617}}\end{smallmatrix}\!\right], \\
M_{21} &= \left[\!\begin{smallmatrix}1& -2& 0\\ \tfrac{-4}{\sqrt[4]{617}}& \tfrac{4}{\sqrt[4]{617}}& \tfrac{-4}{\sqrt{617}}\end{smallmatrix}\!\right], \text{ and} \\
M_{22} &= \left[\!\begin{smallmatrix}0 \\ \tfrac{\sqrt{\sqrt{617}+4}}{\sqrt[4]{617}} \end{smallmatrix}\!\right] \left[\!\begin{smallmatrix}0 & \tfrac{\sqrt{\sqrt{617}+4}}{\sqrt[4]{617}}\end{smallmatrix}\!\right], 
\end{align*}
where the factorization for $M_{11}$ is obtained by (i) finding a matrix $T \in \mathbb{R}^{3 \times 2}$ such that the columns of $M_{11}T$ span the column space of $M_{11}$; then (ii) computing a Cholesky decomposition for $T^{T}M_{11}T$. 
The factorization of $M_{22}$ can be found similarly. Finally, using \citep[Sections 9.2 and 9.4]{AndVong}, we find that $\mathcal{B}$ is realized by the RLCT network in Fig.\ \ref{fig:ex1}.

Our final example considers the behavior $\mathcal{B}$ in (\ref{eq:bgd}), with
\begin{equation*}
P(\xi) = \left[\!\begin{smallmatrix}1 & 1& -1\\ 0& \xi& 0\\ \xi + 1& 1& 0\end{smallmatrix}\!\right], \hspace{0.1cm} Q(\xi) = \left[\!\begin{smallmatrix}0& 0& 0\\ -1& \xi^2 + 1& \xi^2\\ \xi + 2& \xi - 1& 2\xi + 1\end{smallmatrix}\!\right],
\end{equation*}
for which $Q$ is singular. We use this example to illustrate both the proof of Theorem \ref{thm:prbtp1} and the inductive procedure described in Lemma \ref{thm:ppst}. Again, the realization procedure works in the general case. In addition to the previously listed algorithms, it relies on the computation of an upper echelon form for a polynomial matrix.

First, following the proof of Theorem \ref{thm:prbtp1}, we obtain matrices $T = [T_{1} \hspace{0.15cm} T_{2}]^{T}$ and $\hat{Y} \in \mathbb{R}^{3 \times 3}[\xi]$, where
\begin{equation*}
T_{1} = \left[\!\begin{smallmatrix}1& 0\\0& 1\\0& 0\end{smallmatrix}\!\right], \hspace{0.05cm} T_{2} = \left[\!\begin{smallmatrix}{-}1 \\{-}1 \\1\end{smallmatrix}\!\right], \hspace{0.05cm} \hat{Y}(\xi) = \left[\!\begin{smallmatrix}0& 0 & {-}1 \\ {-}\xi^2 {-} 1& {-}\tfrac{1}{2}\xi {-} \tfrac{1}{2}& 0\\ 1{-} \xi& {-}\tfrac{1}{2}& 0\end{smallmatrix}\!\right], 
\end{equation*}
and we find that
\begin{align*}
PT^{-1} &= \hat{Y}\left[\!\begin{smallmatrix}P_{1}& 0\\ 0& 1\end{smallmatrix}\!\right], \text{ and } QT^T = \hat{Y}\left[\!\begin{smallmatrix}Q_{1}& 0\\ 0& 0\end{smallmatrix}\!\right], \text{ where}\\
P_{1}(\xi) &{=} \left[\!\begin{smallmatrix}\tfrac{1}{2}\xi^2 {+} \xi {+} \tfrac{1}{2}& \tfrac{1}{2}\\ {-}\xi^3 {-} \xi^2 {-} \xi {-} 1& {-}\xi {-} 1\end{smallmatrix}\!\right], \hspace{0.05cm} Q_{1}(\xi) {=} \left[\!\begin{smallmatrix} \tfrac{1}{2} \xi^2 {+} \tfrac{3}{2}\xi {+} \tfrac{3}{2}& -1\\ -\xi^3 {-} 2\xi^2 {-} 2\xi {-} 1& 0\end{smallmatrix}\!\right].
\end{align*}
Here, for any given $\lambda > 0$, then $T \in \mathbb{R}^{3 \times 3}$ is a nonsingular matrix such that $T_{2}$ is a basis for the right nullspace of $Q(\lambda)$. Also, $\hat{Y}$ and $Q_{1}$ are obtained by computing an upper echelon form for $QT^{T}$. 
It then follows from the proof of Theorem \ref{thm:prbtp1} that $\mathcal{B}$ is realized by a network of the form shown in Fig.\ \ref{fig:rlct_gen}, where $N_{a,1}$ is a short circuit, and $N_{a,2}$ is a network whose driving-point behavior is the set of solutions to $P_{1}(\tfrac{d}{dt})\mathbf{i}_{1} = Q_{1}(\tfrac{d}{dt})\mathbf{v}_{1}$.

Next, note that $\lim_{\xi \rightarrow \infty}((Q_{1}^{-1}P_{1})(\xi)) = \text{diag}(1 \hspace{0.15cm} 0)$, which is singular. Thus, $(P_{1}, Q_{1})$ satisfies conditions \ref{nl:ip1}--\ref{nl:ip3} on p.\ \pageref{nl:ip1}, but not condition \ref{nl:ip4}. Then, following Lemma \ref{lem:lrz}, we find that $\lim_{\xi \rightarrow \infty}(\tfrac{1}{\xi}(P_{1}^{-1}Q_{1})(\xi)) = \text{diag}(0 \hspace{0.15cm} 1)$, and accordingly we let $K = 1$, $Q_{2} = P_{1}$, and
\begin{equation*}
P_{2}(\xi) = Q_{1}(\xi) - P_{1}(\xi)\left[\!\begin{smallmatrix}0& 0\\0& \xi\end{smallmatrix}\!\right] = \left[\!\begin{smallmatrix}\tfrac{1}{2}\xi^2 {+} \tfrac{3}{2}\xi {+} \tfrac{3}{2}& {-}1 {-} \tfrac{1}{2}\xi \\ {-}\xi^3 {-} 2\xi^2 {-} 2\xi {-} 1& \xi^2 {+} \xi\end{smallmatrix}\!\right].
\end{equation*}
In this case, $\lim_{\xi \rightarrow \infty}((Q_{2}^{-1}P_{2})(\xi)) = \text{diag}(1 \hspace{0.15cm} 0)$, so again we apply Lemma \ref{lem:lrz}. Here, $\lim_{\xi \rightarrow \infty}(\tfrac{1}{\xi}(P_{2}^{-1}Q_{2})(\xi)) = \text{diag}(0 \hspace{0.15cm} 1)$, and we let $K = 1$, $Q_{3} = P_{2}$, and
\begin{equation*}
P_{3}(\xi) = Q_{2}(\xi) - P_{2}(\xi)\left[\!\begin{smallmatrix}0& 0\\0& \xi\end{smallmatrix}\!\right] = \left[\!\begin{smallmatrix}\tfrac{1}{2}\xi^2 {+} \xi {+} \tfrac{1}{2}& \tfrac{1}{2}\xi^2 {+} \xi {+} \tfrac{1}{2} \\ {-}\xi^3 {-} \xi^2 {-} \xi {-} 1& {-}\xi^3 {-} \xi^2 {-} \xi {-} 1\end{smallmatrix}\!\right].
\end{equation*}
Next, we find that $(P_{3},Q_{3})$ satisfies condition  \ref{nl:ip1} on p.\ \pageref{nl:ip1}, but not condition \ref{nl:ip3}. Thus, following Lemma \ref{lem:gprl2}, we let $\hat{T} = [\hat{T}_{1} \hspace{0.15cm} \hat{T}_{2}]$ and $W \in \mathbb{R}^{2 \times 2}[\xi]$, where
\begin{equation*}
\hat{T}_{1} = \left[\!\begin{smallmatrix}1 \\ 0\end{smallmatrix}\!\right], \hspace{0.05cm} \hat{T}_{2} = \left[\!\begin{smallmatrix}1 \\ -1\end{smallmatrix}\!\right] \text{ and } W(\xi) = \left[\!\begin{smallmatrix}1 - \xi& -\tfrac{1}{2}\\ -2\xi^2 - 2& -1 - \xi\end{smallmatrix}\!\right],
\end{equation*} 
and we find that
\begin{equation*}
P_{3}\hat{T} = W^{-1}\left[\!\begin{smallmatrix}P_{4}& 0\\0& 0\end{smallmatrix}\!\right], \hspace{0.1cm} Q_{3}(\hat{T}^{-1})^{T} = W^{-1}\left[\!\begin{smallmatrix}Q_{4}& 1\\ 0 & -2\end{smallmatrix}\!\right],
\end{equation*}
where $P_{4}(\xi) = Q_{4}(\xi) = \xi + 1$. In this case, $\hat{T} \in \mathbb{R}^{2 \times 2}$ is a nonsingular matrix such that $\hat{T}_{2}$ is a basis for the right nullspace of $P_{3}$, and the matrices $W$ and $Q_{4}$ are obtained from the upper echelon form for $Q_{3}(\hat{T}^{-1})^{T}$. 
It can then be verified that $\mathcal{B}$ is realized by a network of the form of Fig.\ \ref{fig:ex2}. Here, 
a network realization for the set of solutions to the differential equation $(\tfrac{d}{dt}+1)i_{e} = (\tfrac{d}{dt}+1)v_{e}$ can be obtained by the method outlined in the first example. 

\section{Conclusions}
This paper developed a theory of reciprocal systems which does not assume controllability. Necessary and sufficient algebraic conditions were established for a system to be reciprocal, both in terms of the high order differential equations describing the system, and in terms of a state-space realization. Analogous results were obtained for systems that are both passive and reciprocal. Notably, we answered the first open problem in \citep{camwb} by proving that a behavior is realizable as the driving-point behavior of an RLCT network if and only if it is passive and reciprocal.

%

\appendix
\section{The elimination theorem}
\label{app:b}
Let $\hat{\mathcal{B}} = \lbrace (\mathbf{w}_{1}, \mathbf{w}_{2}) \in \mathcal{L}_{1}^{\text{loc}}\left(\mathbb{R}, \mathbb{R}^{n_{1}}\right) \times \mathcal{L}_{1}^{\text{loc}}\left(\mathbb{R}, \mathbb{R}^{n_{2}}\right) \mid \hat{R}(\tfrac{d}{dt})\text{col}(\mathbf{w}_{1} \hspace{0.15cm} \mathbf{w}_{2})\rbrace$. From \citep[Theorem 6.2.6]{JWIMTSC}, there exists a unimodular $U$ with
\begin{equation}
U\hat{R} = \begin{bmatrix}R_{1}& 0\\ R_{2}& M_{2}\end{bmatrix}, \label{eq:epf}
\end{equation}
where the rightmost matrix is partitioned compatibly with $\text{col}(\mathbf{w}_{1} \hspace{0.15cm} \mathbf{w}_{2})$, and $M_{2}$ has full row rank. Then, from \citep[Theorem 2.5.4]{JWIMTSC}, $\hat{\mathcal{B}}$ is the set of locally integrable solutions to $R_{1}(\tfrac{d}{dt})\mathbf{w}_{1} = 0$ and $M_{2}(\tfrac{d}{dt})\mathbf{w}_{2} = -R_{2}(\tfrac{d}{dt})\mathbf{w}_{1}$. Now, let $\mathcal{B} := \lbrace \mathbf{w}_{1} \in \mathcal{L}_{1}^{\text{loc}}\left(\mathbb{R}, \mathbb{R}^{n_{1}}\right) \mid R_{1}(\tfrac{d}{dt})\mathbf{w}_{1} = 0\rbrace$. Since $M_{2}$ has full row rank, then it is easily shown that for any $\mathbf{w}_{1} \in \mathcal{D}_{+}\left(\mathbb{R}, \mathbb{R}^{n_{1}}\right)$ there exists  $\mathbf{w}_{2} \in \mathcal{D}_{+}\left(\mathbb{R}, \mathbb{R}^{n_{2}}\right)$ such that $M_{2}(\tfrac{d}{dt})\mathbf{w}_{2} = -R_{2}(\tfrac{d}{dt})\mathbf{w}_{1}$, whence $(\hat{\mathcal{B}} \cap (\mathcal{D}_{+}\left(\mathbb{R}, \mathbb{R}^{n_{1}}\right) \times \mathcal{D}_{+}\left(\mathbb{R}, \mathbb{R}^{n_{2}}\right)))^{(\mathbf{w}_{1})} = \mathcal{B} \cap \mathcal{D}_{+}\left(\mathbb{R}, \mathbb{R}^{n_{1}}\right)$. But it may not be the case that $\hat{\mathcal{B}}^{(\mathbf{w}_{1})} = \mathcal{B}$ \citep[see, e.g.,][Example 2.1]{JWPPE}. If $\hat{\mathcal{B}}^{(\mathbf{w}_{1})} = \mathcal{B}$, then $\mathbf{w}_{2}$ is called \emph{properly eliminable} \citep{JWPPE}. From \citep[Example 3.1]{JWPPE}, if $\mathcal{B}_{s}$ is as in (\ref{eq:bhssr}), then $\mathbf{x}$ is properly eliminable. Also, the internal currents and voltages in any given RLCT network are properly eliminable  \citep[Section 6]{HUGIFAC}.

Finally, if $\mathcal{B}$ is as in (\ref{eq:bd}) and $T \in \mathbb{R}^{q \times q}$ is a nonsingular real matrix, then 
it is easily shown that $\mathcal{B}^{(T\mathbf{w})} = \lbrace \mathbf{z} \in \mathcal{L}_{1}^{\text{loc}}\left(\mathbb{R}, \mathbb{R}^{q}\right) \mid (RT^{-1})(\tfrac{d}{dt})\mathbf{z} {=} 0 \rbrace$.

\section{The passive and reciprocal behavior theorem, supplementary lemmas}
\label{sec:prbtsl}
In this appendix, we present four supplementary lemmas used to prove the results in Sections \ref{sec:prb} and \ref{sec:raps}. In the first two lemmas, for any given symmetric $K \in \mathbb{R}^{d \times d}$, we let
\begin{align}
\hspace*{-0.4cm}&\Upsilon(K) := -KA^{T} {-} AK {-} (KC^{T}{-}B)(D{+}D^{T})^{-1}(CK{-}B^{T}), \nonumber \\
\hspace*{-0.4cm}&\text{and } A_{\Upsilon}(K) := A^{T} {-} C^{T}(D{+}D^{T})^{-1}(B^{T}{-}CK). \label{eq:upapsid}
\end{align}

\begin{lemapp}
\label{thm:obspds}
Let $\mathcal{B}_{s}$ be as in (\ref{eq:bhssr}), and let $\hat{\mathcal{B}} := \mathcal{B}_{s}^{(\mathbf{u},\mathbf{y})}$ be passive, $(C, A)$ be observable, $D + D^{T} > 0$, and $\Upsilon(K), A_{\Upsilon}(K)$ be as in (\ref{eq:upapsid}). Then there exists $K \in \mathbb{R}^{d \times d}$ such that $K > 0$, $\Upsilon(K) \geq 0$, and $\text{spec}(A_{\Upsilon}(K)) \in \overbar{\mathbb{C}}_{-}$.
\end{lemapp}

\begin{pf}
Since $(C,A)$ is observable then there exists $X \in \mathbb{R}^{d \times d}$ such that $X > 0$ and $-A^{T}X - XA - (C^{T} - XB)(D+D^{T})^{-1}(C - B^{T}X) = 0$ \citep[see][Theorem 13]{THTPLSNA}. Now, let $K_{1} := X^{-1} \in \mathbb{R}^{d \times d}$, so $K_{1} > 0$ and $\Upsilon(K_{1}) = 0$. Thus, from \citep[Theorems 10 and 11]{THAS}, there exists $K_{-} \geq 0$ such that $\Upsilon(K_{-}) = 0$, $\text{spec}(A_{\Upsilon}(K_{-})) \in \overbar{\mathbb{C}}_{-}$, and $K_{-} \leq K_{1}$ (here, the available energy $S_{a}$ for the system $\tfrac{d\mathbf{x}}{dt} = A^{T}\mathbf{x} - C^{T}\mathbf{u}$, $\mathbf{y} = -B^{T}\mathbf{x} + D^{T} \mathbf{u}$ satisfies $S_{a}(\mathbf{x}_{0}) = \mathbf{x}_{0}^{T}K_{-}\mathbf{x}_{0}$ for all $\mathbf{x}_{0} \in \mathbb{R}^{d}$). Now, let $\epsilon$ be a fixed but arbitrary real number in the interval $0 < \epsilon < 1$, and let $K_{\epsilon} := (1- \epsilon) K_{-} + \epsilon K_{1}$. Since $\epsilon K_{1} > 0$ and $(1- \epsilon) K_{-} \geq 0$, then $K_{\epsilon} > 0$. Also, 
\begin{align*}
&\Upsilon(K_{\epsilon}) = (1{-} \epsilon)\Upsilon(K_{-}) + \epsilon \Upsilon(K_{1}) \\ &\hspace{0.6cm} + \epsilon(1{-}\epsilon)(K_{-}{-}K_{1})C^{T}(D{+}D^{T})^{-1}C(K_{-}{-}K_{1}) \geq 0,
\end{align*}
and so $\Upsilon(K_{\epsilon}) \geq 0$. To complete the proof of the present theorem, we will show that there exists $0 < \alpha < 1$ such that $\text{spec}(A_{\Upsilon}(K_{\epsilon})) \in \overbar{\mathbb{C}}_{-}$ for all $0 < \epsilon \leq \alpha$. To see this, note that $Z := K_{1} - K_{-}$ satisfies $Z \geq 0$ and 
\begin{equation*}
-Z A_{\Upsilon}(K_{-}) - A_{\Upsilon}(K_{-})^{T}Z = ZC^{T}(D + D^{T})^{-1}CZ.
\end{equation*}
Next, let $T \in \mathbb{R}^{d \times d}$ be nonsingular with $T A_{\Upsilon}(K_{-})^{T}T^{-1} = \text{diag}(A_{1} \hspace{0.15cm} A_{2})$ where $\text{spec}(A_{1}) \in \mathbb{C}_{-}$ and $\text{spec}(A_{2}) \in j\mathbb{R}$ (here, the rows of $T_{1}$ span the stable left eigenspace of $A_{\Upsilon}(K_{-})^{T}$), and partition $T$ compatibly as $T = \text{col}(T_{1} \hspace{0.15cm} T_{2})$. Then the row space of $T_{2}$ is spanned by the left Jordan chains corresponding to the imaginary axis eigenvalues of $A_{\Upsilon}(K_{-})$. Consider one such Jordan chain: 
\begin{align*}
\mathbf{z}_{1}^{T}A_{\Upsilon}(K_{-})^{T} &= j \omega \mathbf{z}_{1}^{T}, \text{ and} \\
\mathbf{z}_{k}^{T}A_{\Upsilon}(K_{-})^{T} &= j \omega \mathbf{z}_{k}^{T} +  \mathbf{z}_{k-1}^{T} \hspace{0.1cm} (k = 2, 3, \ldots , N).
\end{align*}
Then $A_{\Upsilon}(K_{-})\bar{\mathbf{z}}_{1} = -j\omega \bar{\mathbf{z}}_{1}$ and $A_{\Upsilon}(K_{-})\bar{\mathbf{z}}_{k} = -j\omega \bar{\mathbf{z}}_{k} + \bar{\mathbf{z}}_{k-1}$ ($k = 2, 3, \ldots , N$). Thus, for $k=1$, 
\begin{multline}
\mathbf{z}_{k}^{T}ZC^{T}(D {+} D^{T})^{-1}CZ\bar{\mathbf{z}}_{k} \\
= \mathbf{z}_{k}^{T}(-Z A_{\Upsilon}(K_{-}) {-} A_{\Upsilon}(K_{-})^{T}Z)\bar{\mathbf{z}}_{k} = 0,\label{eq:zkr}
\end{multline}
whence $CZ\bar{\mathbf{z}}_{1} = 0$. This implies that $(-Z A_{\Upsilon}(K_{-}) - A_{\Upsilon}(K_{-})^{T}Z)\bar{\mathbf{z}}_{1} = 0$, so $A_{\Upsilon}(K_{-})^{T}Z\bar{\mathbf{z}}_{1} = -ZA_{\Upsilon}(K_{-}) \bar{\mathbf{z}}_{1} = j\omega Z\bar{\mathbf{z}}_{1}$. It follows that $C (Z\bar{\mathbf{z}}_{1}) = 0$ and  $A(Z\bar{\mathbf{z}}_{1}) = A_{\Upsilon}(K_{-})^{T}(Z\bar{\mathbf{z}}_{1}) = j\omega Z\bar{\mathbf{z}}_{1}$, and so $Z\bar{\mathbf{z}}_{1} = 0$ since $(C,A)$ is observable. Next, note that (\ref{eq:zkr}) holds for $k=2$, and similar to before we find that $Z\bar{\mathbf{z}}_{2} = 0$. Proceeding by induction, we obtain $Z\bar{\mathbf{z}}_{k} = 0$, whence $\mathbf{z}_{k}^{T}Z = 0$ ($k = 1, 2, \ldots , N$). Since the vectors $\mathbf{z}_{1} \ldots \mathbf{z}_{N}$ span the row space of $T_{2}$, then $T_{2}Z = 0$. Thus, by partitioning $\hat{T} := T^{-1}$ compatibly with $T$ as $\hat{T} = [\hat{T}_{1} \hspace{0.15cm} \hat{T}_{2}]$, noting that $A_{\Upsilon}(K_{\epsilon})^{T} = A_{\Upsilon}(K_{-})^{T} + \epsilon ZC^{T}(D{+}D^{T})^{-1}C$, and letting $\hat{A}_{12} := \epsilon T_{1}ZC^{T}(D{+}D^{T})^{-1}C\hat{T}_{2}$, we find that
\begin{equation*}
TA_{\Upsilon}(K_{\epsilon})^{T}T^{-1} = \begin{bmatrix}A_{1} {+} \epsilon T_{1}ZC^{T}(D{+}D^{T})^{-1}C\hat{T}_{1}& \hat{A}_{12} \\0 & A_{2}\end{bmatrix}.
\end{equation*}
Thus, $\text{spec}(A_{\Upsilon}(K_{\epsilon})) = \text{spec}(A_{1} {+} \epsilon T_{1}ZC^{T}(D{+}D^{T})^{-1}C\hat{T}_{1})$ $\cup \text{spec}(A_{2})$. Since $\text{spec}(A_{1}) \in \mathbb{C}_{-}$, then there exists a $0 < \alpha < 1$ such that $\text{spec}(A_{1} {+} \epsilon T_{1}ZC^{T}(D{+}D^{T})^{-1}C\hat{T}_{1}) \in \overbar{\mathbb{C}}_{-}$ for all $0 < \epsilon \leq \alpha$. For any such $\epsilon$, then $K := K_{\epsilon}$ satisfies the conditions of the present theorem statement. \qed
\end{pf}

\begin{lemapp}
\label{lem:pdrr}
Let $\mathcal{B}_{s}$ be as in (\ref{eq:bhssr}), and let $\hat{\mathcal{B}} := \mathcal{B}_{s}^{(\mathbf{u},\mathbf{y})}$ be passive, $(C, A)$ be detectable (i.e., $\text{col}(C \hspace{0.15cm} \lambda I - A)$ has full column rank for all $\lambda \in \overbar{\mathbb{C}}_{+}$), $D + D^{T} > 0$, and $\Upsilon(K)$ be as in (\ref{eq:upapsid}). Then there exists $K \in \mathbb{R}^{d \times d}$ such that $K > 0$ and $\Upsilon(K) \geq 0$.
\end{lemapp}

\begin{pf}
By the observer staircase form \citep[see][note D2]{THTPLSNA}, there exists a $T \in \mathbb{R}^{d \times d}$ such that
\begin{equation*}
TAT^{-1} = \begin{bmatrix}\tilde{A}_{11}& 0\\ \tilde{A}_{21}& \tilde{A}_{22}\end{bmatrix}, \hspace{0.1cm} TB = \begin{bmatrix}\tilde{B}_{1}\\ \tilde{B}_{2}\end{bmatrix}, \hspace{0.1cm} CT^{-1} = \begin{bmatrix}\tilde{C}_{1}& 0\end{bmatrix},
\end{equation*}
with $(\tilde{C}_{1}, \tilde{A}_{11})$ observable. 
As $(C, A)$ is detectable, then it is easily shown that $\text{spec}(\tilde{A}_{22}) {\in} \mathbb{C}_{-}$. Now, let 
\begin{align*}
&\tilde{\mathcal{B}}_{s} = \lbrace (\mathbf{u}, \mathbf{y}, \tilde{\mathbf{x}}) {\in} \mathcal{L}_{1}^{\text{loc}}\left(\mathbb{R}, \mathbb{R}^{n}\right) {\times} \mathcal{L}_{1}^{\text{loc}}\left(\mathbb{R}, \mathbb{R}^{n}\right) {\times} \mathcal{L}_{1}^{\text{loc}}\left(\mathbb{R}, \mathbb{R}^{\tilde{d}}\right) \mid \\
& \hspace{2cm} \tfrac{d\tilde{\mathbf{x}}}{dt} = \tilde{A}_{11}\tilde{\mathbf{x}} + \tilde{B}_{1}\mathbf{u} \text{ and } \mathbf{y} = \tilde{C}_{1}\tilde{\mathbf{x}} + D\mathbf{u} \rbrace, \\
&\tilde{\Upsilon}(\tilde{K}) := -\tilde{K}\tilde{A}_{11}^{T} - \tilde{A}_{11}\tilde{K}\\
&\hspace{2cm} - (\tilde{K}\tilde{C}_{1}^{T} - \tilde{B}_{1})(D + D^{T})^{-1}(\tilde{C}_{1}\tilde{K} - \tilde{B}_{1}^{T}), \\
&\text{and } \tilde{A}_{\tilde{\Upsilon}}(\tilde{K}) = \tilde{A}_{11}^{T} - \tilde{C}_{1}^{T}(D+D^{T})^{-1}(\tilde{B}_{1}^{T} - \tilde{C}_{1}\tilde{K}).
\end{align*}
It follows from \citep[Note D3]{THTPLSNA} that $\tilde{\mathcal{B}}_{s}^{(\mathbf{u}, \mathbf{y})} = \mathcal{B}_{s}^{(\mathbf{u}, \mathbf{y})}$, which is passive, whence from Lemma \ref{thm:obspds} there exists $K_{11} \in \mathbb{R}^{\tilde{d} \times \tilde{d}}$ such that $K_{11} > 0$, $\tilde{\Upsilon}(K_{11}) \geq 0$, and $\text{spec}(\tilde{A}_{\tilde{\Upsilon}}(K_{11})) \in \overbar{\mathbb{C}}_{-}$. Since, in addition, $\text{spec}(\tilde{A}_{22}) \in \mathbb{C}_{-}$, then by \citep[Theorem 3.7.4]{AndVong} there exists a unique real $K_{12}$ which satisfies the Sylvester equation
\begin{align*}
&\tilde{A}_{22}K_{12}^{T} + K_{12}^{T}\tilde{A}_{\tilde{\Upsilon}}(K_{11}) \\ &\hspace{0.5cm}= -\tilde{A}_{21}K_{11} - \tilde{B}_{2}(D + D^{T})^{-1}(\tilde{B}_{1}^{T} - \tilde{C}_{1}K_{11});
\end{align*}
and there exists a non-unique $\nabla > 0$ which satisfies
\begin{align*}
&\Psi(\nabla) := -\nabla \tilde{A}_{22}^{T} -\tilde{A}_{22}\nabla - K_{12}^{T}K_{11}^{-1}\tilde{\Upsilon}(K_{11})K_{11}^{-1}K_{12} \\
&{-}(\tilde{B}_{2}{-} K_{12}^{T}K_{11}^{-1}\tilde{B}_{1})(D {+} D^{T})^{-1}(\tilde{B}_{2}{-} K_{12}^{T}K_{11}^{-1}\tilde{B}_{1})^{T}  \geq 0.
\end{align*}
It can then be verified that 
\begin{align*}
K {:=} T^{-1}\begin{bmatrix}I& 0\\K_{12}^{T}K_{11}^{-1}& I\end{bmatrix}\!\begin{bmatrix}K_{11}&0\\0& \nabla\end{bmatrix}\!\begin{bmatrix}I& K_{11}^{-1}K_{12}\\ 0& I\end{bmatrix}(T^{-1})^{T} {>} 0, \\
\text{and } \Upsilon(K) = T^{-1} \begin{bmatrix}\tilde{\Upsilon}(K_{11})& 0\\0& \Psi(\nabla)\end{bmatrix}(T^{-1})^{T} \geq 0. \hfill \qed
\end{align*}
\end{pf}
\begin{remapp}
\textnormal{It is easily shown that $K$ in Lemma \ref{lem:pdrr} satisfies $\text{spec}(A_{\Upsilon}(K)) {=} \text{spec}(A_{\tilde{\Upsilon}}(K_{11})) {\cup} \text{spec}(A_{22}) {\in} \overbar{\mathbb{C}}_{-}$.}
\end{remapp}

The final two lemmas concern the decomposition in the proof of Lemma \ref{thm:ppst}. We refer to that proof for the definition of statements \ref{nl:ip1}--\ref{nl:ip4} and \ref{nl:csn1}--\ref{nl:csn2}.
\begin{lemapp}
\label{lem:gprl2}
Let $P_{k{-}1}, Q_{k{-}1}$ satisfy \ref{nl:ip1} for $i = k{-}1$, and let $n_{k} := \text{normalrank}(P_{k{-}1})$, $m_{k} := n_{k{-}1} - n_{k}$, 
and $r_{k} := \text{rank}(\lim_{\xi \rightarrow \infty}(Q_{k{-}1}^{-1}P_{k{-}1}(\xi)))$. The following hold.
\begin{enumerate}[label=\arabic*., ref=\arabic*, leftmargin=0.5cm]
\item There exists a nonsingular $T \in \mathbb{R}^{n_{k{-}1} \times n_{k{-}1}}$; unimodular $W \in \mathbb{R}^{n_{k{-}1} \times n_{k{-}1}}[\xi]$ and $\tilde{Q}_{22} \in \mathbb{R}^{m_{k} {\times} m_{k}}[\xi]$; $\tilde{Q}_{12} \in \mathbb{R}^{n_{k} {\times} m_{k}}[\xi]$; and $P_{k}, Q_{k}$ satisfying \ref{nl:ip1} and \ref{nl:ip3} for $i {=} k$, with\label{nl:fns1}
\begin{equation}
\hspace*{-0.6cm} WP_{k{-}1}T {=} \begin{bmatrix}P_{k}& 0 \\0 & 0\end{bmatrix}\!, WQ_{k{-}1}(T^{-1})^{T} {=} \begin{bmatrix}Q_{k}& \tilde{Q}_{12}\\0& \tilde{Q}_{22}\end{bmatrix}\!.\label{eq:tmfoz}
\end{equation}
\item  Let $A_{k}, B_{k}, C_{k}, D_{k}$ satisfy \ref{nl:csn1} for $i = k$; and let $A_{k{-}1} {:=} A_{k}$, $B_{k{-}1} {:=} [B_{k} \hspace{0.15cm} 0]T^{-1}$, $C_{k{-}1} {:=} (T^{-1})^{T}\text{col}(C_{k} \hspace{0.15cm} 0)$, and $D_{k{-}1} {:=} (T^{-1})^{T}\text{diag}(D_{k} \hspace{0.15cm} 0)T^{-1}$. Then:\label{nl:fnsa2}
\begin{enumerate}
\item  \ref{nl:csn1} holds for $i=k{-}1$.\label{nl:fns2}
\item Let $X_{k}$ satisfy \ref{nl:csn2-} for $i = k$; and let $X_{k{-}1} := X_{k}$. Then \ref{nl:csn2-} holds for $i = k{-}1$.\label{nl:fns4}
\item Let $S_{k}$ satisfy \ref{nl:csn2} for $i = k$; and let $S_{k{-}1} := S_{k}$. Then \ref{nl:csn2} holds for $i = k{-}1$.\label{nl:fns5}
\end{enumerate}
\end{enumerate}
\end{lemapp}

\begin{pf}
Condition \ref{nl:fns1} follows from \cite[Lemma D.3, condition 1]{THAS}, noting that $T^{T}Q_{k{-}1}^{-1}P_{k{-}1}T = \text{diag}(Q_{k}^{-1}P_{k} \hspace{0.15cm} 0)$, so $Q_{k}^{-1}P_{k}$ is symmetric since $Q_{k{-}1}^{-1}P_{k{-}1}$ is. To see condition \ref{nl:fns2}, we let $\mathcal{A}_{k}, Y_{k}, Z_{k}, U_{k}, V_{k}, E_{k}, F_{k}$ and $G_{k}$ be as in \ref{nl:csn1} for $i = k$. Following \cite[Lemma D.3, proof of condition 2]{THAS}, we let
\begin{equation*}
\begin{bmatrix}Y_{k{-}1} & Z_{k{-}1}\\ U_{k{-}1} & V_{k{-}1}\end{bmatrix} := \begin{bmatrix}W^{-1}& 0\\ 0& I\end{bmatrix} \left[\begin{array}{cc|c}Y_{k}& \tilde{Q}_{12} & Z_{k} \\ 0 & \tilde{Q}_{22} & 0 \\ \hline \rule{0pt}{1.1\normalbaselineskip}
 U_{k} & 0 & V_{k}\end{array}\right] \begin{bmatrix}T^{T}& 0\\ 0& I\end{bmatrix}.
\end{equation*}
It can be verified that each of the above matrices is unimodular. Also, with $\mathcal{A}_{k{-}1}(\xi) := \xi I - A_{k{-}1}$, $E_{k{-}1} := [E_{k} \hspace{0.15cm} 0]T^{-1}$, $F_{k{-}1} := [F_{k} \hspace{0.15cm} 0]T^{T}$, and $G_{k{-}1} := G_{k}$, it can be verified that \ref{nl:csn1} holds for $i = k-1$.

The proof of condition \ref{nl:fns4} follows from \cite[Lemma D.3, proof of condition 2(c)]{THAS}: with $R := \text{diag}(I \hspace{0.15cm} T^{-1})$, then $\Omega_{k{-}1}(X_{k{-}1}) = R^{T}\text{diag}(\Omega_{k}(X_{k}) \hspace{0.15cm} 0)R$. Finally, condition \ref{nl:fns5} is straightforward to check. \qed
\end{pf}

\begin{rem}
\textnormal{
With $P_{k}, Q_{k}, P_{k-1}$ and $Q_{k-1}$ as in the above lemma, then the driving-point behavior $P_{k-1}(\tfrac{d}{dt})\mathbf{i} = Q_{k-1}(\tfrac{d}{dt})\mathbf{v}$ can be realized by a transformer terminated on a network with driving-point behavior $P_{k}(\tfrac{d}{dt})\mathbf{\hat{i}} = Q_{k}(\tfrac{d}{dt})\mathbf{\hat{v}}$ (see the final example in Section \ref{sec:ex}).
}
\end{rem}

\begin{lemapp}
\label{lem:lrz}
Let  $P_{k{-}1}, Q_{k{-}1}$ satisfy \ref{nl:ip1}--\ref{nl:ip3} for $i {=} k{-}1$, with $m_{k} :=  n_{k{-}1} - r_{k{-}1} > 0$. The following hold.
\begin{enumerate}[label=\arabic*., ref=\arabic*, leftmargin=0.5cm]
\item There exists $0 < K \in \mathbb{R}^{m_{k} \times m_{k}}$ such that $\text{diag}(0 \hspace{0.15cm} K) = \lim_{\xi \rightarrow \infty}(\tfrac{1}{\xi}P_{k{-}1}^{-1}Q_{k{-}1}(\xi))$. 
\label{nl:lrz1} 
\item Let $P_{k}(\xi) := Q_{k{-}1}(\xi) - P_{k{-}1}(\xi)\text{diag}(0 \hspace{0.15cm} K\xi)$, and $Q_{k} := P_{k{-}1}$. Then \ref{nl:ip1} holds for $i = k$; $\deg{(\det{(Q_{k})})} < \deg{(\det{(Q_{k{-}1})})}$; and there exist $\hat{D}_{12} \in \mathbb{R}^{r_{k{-}1}{\times}m_{k}}, \hat{D}_{21} \in \mathbb{R}^{m_{k}{\times}r_{k{-}1}}, \hat{D}_{22} \in  \mathbb{R}^{m_{k}{\times}m_{k}}$ such that\label{nl:lrz2}
\begin{equation}
\lim_{\xi \rightarrow \infty}(Q_{k}^{-1}P_{k}(\xi)) =: D_{k} = \begin{bmatrix}I_{r_{k{-}1}} & \hat{D}_{12}\\ \hat{D}_{21}& \hat{D}_{22}\end{bmatrix}.\label{eq:digf} 
\end{equation}
\item Let $A_{k}, B_{k}, C_{k}, D_{k}$ satisfy \ref{nl:csn1} for $i = k$; partition $B_{k}, C_{k}$ compatibly with $D_{k}$ as $B_{k} = [\hat{B}_{1} \hspace{0.15cm} \hat{B}_{2}]$, $C_{k} = \text{col}(\hat{C}_{1} \hspace{0.15cm} \hat{C}_{2})$; and let 
\begin{align*}
&A_{k{-}1} := \begin{bmatrix} A_{k} - \hat{B}_{1} \hat{C}_{1} & \hat{B}_{2} K^{-1} - \hat{B}_{1} \hat{D}_{12} K^{-1}\\  \hat{D}_{21} \hat{C}_{1} - \hat{C}_{2} & \hat{D}_{21} \hat{D}_{12} K^{-1} - \hat{D}_{22} K^{-1}\end{bmatrix}, \\ 
&B_{k{-}1} := \begin{bmatrix} \hat{B}_{1} & 0 \\  - \hat{D}_{21} & I\end{bmatrix}, \text{ and } C_{k{-}1} := \begin{bmatrix} - \hat{C}_{1} &  - \hat{D}_{12} K^{-1} \\ 0 & K^{-1}\end{bmatrix}. 
\end{align*}
Then:\label{nl:lrza3}
\begin{enumerate}
\item \ref{nl:csn1} holds for $i = k{-}1$.\label{nl:lrz4}
\item Let $X_{k}$ satisfy \ref{nl:csn2-} for $i = k$; and let $X_{k{-}1} := \text{diag}(X_{k} \hspace{0.15cm} K^{-1})$. Then \ref{nl:csn2-} holds for $i = k{-}1$.\label{nl:lrz6}
\item Let $S_{k}$ satisfy \ref{nl:csn2} for $i = k$; and let $S_{k{-}1} := \text{diag}({-}S_{k} \hspace{0.15cm} K^{-1})$. Then \ref{nl:csn2} holds for $i = k{-}1$.\label{nl:lrz7}
\end{enumerate}
\end{enumerate}
\end{lemapp}

\begin{pf}
First, note that $Q_{k{-}1}^{-1}P_{k{-}1} = P_{k{-}1}^{T}(Q_{k{-}1}^{-1})^{T}$ implies that $P_{k{-}1}Q_{k{-}1}^{T} = Q_{k{-}1}P_{k{-}1}^{T}$, and hence $P_{k{-}1}^{-1}Q_{k{-}1} = Q_{k{-}1}^{T}(P_{k{-}1}^{T})^{-1}$. Conditions \ref{nl:lrz1} and \ref{nl:lrz2} then follow from \cite[Lemma D.4, conditions 1 and 2]{THAS}, as $Q_{k}^{-1}P_{k}(\xi) = P_{k{-}1}^{-1}Q_{k{-}1}(\xi) - \text{diag}(0 \hspace{0.15cm} K\xi)$, so $Q_{k}^{-1}P_{k}$ 
is symmetric since $P_{k{-}1}^{-1}Q_{k{-}1}$ and $\text{diag}(0 \hspace{0.15cm} K\xi)$ are. For condition \ref{nl:lrz4}, we let $\mathcal{A}_{k}, Y_{k}, Z_{k}, U_{k}, V_{k}$, $E_{k}, F_{k}$ and $G_{k}$ be as in \ref{nl:csn1} for $i = k$. Following \cite[Lemma D.4, proof of condition 3]{THAS}, we partition the two matrices on the left-hand side of \ref{nl:csn1} compatibly as 
\begin{equation}
\left[\!\begin{smallarray}{cc|c}\!\hat{Y}_{11}& \hat{Y}_{12}& \hat{Z}_{1}\\ \hat{Y}_{21}& \hat{Y}_{22}& \hat{Z}_{2}\\ \hline  \rule{0pt}{1.05\normalbaselineskip} \hat{U}_{1}& \hat{U}_{2}& \hat{V}\end{smallarray}\!\right]\! \text{ and } \!\left[\!\begin{smallarray}{cc|c|c}\!\begin{smallarray}{c}-I\\ -\hat{D}_{21}\end{smallarray}\!& \!\begin{smallarray}{c}-\hat{D}_{12}\\ -\hat{D}_{22}\end{smallarray}\!& I& \!\begin{smallarray}{c}-\hat{C}_{1}\\ -\hat{C}_{2}\end{smallarray}\! \\ \hline  \rule{0pt}{1.05\normalbaselineskip} -\hat{B}_{1}& -\hat{B}_{2}& 0& \mathcal{A}_{k} \!\end{smallarray}\!\right]\!, \label{eq:cpf1}
\end{equation}
and we let 
\begin{equation*}
\hspace*{-0.3cm}\begin{bmatrix}Y_{k{-}1}& Z_{k{-}1}\\ U_{k{-}1}& V_{k{-}1}\end{bmatrix}\! = \!\left[\! \begin{smallarray}{cc|cc}\hat{Y}_{11} & \hat{Y}_{12}& \hat{Z}_{1}& 0 \\ \hat{Y}_{21}& \hat{Y}_{22} & \hat{Z}_{2} & 0 \\ \noalign{\hrule} \rule{0pt}{1.05\normalbaselineskip} {-}\hat{U}_{1}& {-}\hat{U}_{2}& {-}\hat{V}& 0\\ 0& I& 0& {-}I\end{smallarray} \!\right]\! \!\left[\!\begin{smallarray}{cc|cc}I& \hat{D}_{12}& 0& 0\\ \hat{D}_{21} &  \hat{D}_{22} {+} K\xi & 0 & I\\ \noalign{\hrule} \rule{0pt}{1.05\normalbaselineskip} \hat{B}_{1} & \hat{B}_{2} & {-}I & 0\\ \hat{D}_{21}& \hat{D}_{22}{+}K(1{+}\xi) & 0 & I\end{smallarray}\!\right]\!.
\end{equation*}
It can be verified that each of the above matrices is unimodular. 
Then, with $E_{k{-}1} := \text{col}(F_{k} \hspace{0.15cm} 0)$, $F_{k{-}1}(\xi) := \text{col}(E_{k}(\xi) \hspace{0.15cm} 0) + \text{col}(\xi\hat{U}_{2}(\xi) \hspace{0.15cm} I)[0 \hspace{0.15cm} K]$, and $G_{k{-}1} {:=} \text{diag}(G_{k}\hspace{0.15cm} I)$, we find that \ref{nl:csn1} holds for $i {=} k{-}1$. 

The proof of condition \ref{nl:lrz6} is identical to \cite[Lemma D.3, proof of condition 3(c)]{THAS}. 
Finally, condition \ref{nl:lrz7} is straightforward to check (noting that $\lim_{\xi \rightarrow \infty}(Q_{k}^{-1}P_{k}(\xi))$ is symmetric, so $\hat{D}_{12} = \hat{D}_{21}^{T}$, and $\hat{D}_{11}$ and $\hat{D}_{22}$ are symmetric). \qed
\end{pf}

\begin{rem}
\textnormal{
With $P_{k}, Q_{k}, P_{k-1}$ and $Q_{k-1}$ as in the above lemma, then the driving-point behavior $P_{k-1}(\tfrac{d}{dt})\mathbf{i} = Q_{k-1}(\tfrac{d}{dt})\mathbf{v}$ can be realized by a parallel connection of networks with driving-point behaviors $\mathbf{\tilde{i}} = \text{diag}(0 \hspace{0.15cm} K\tfrac{d}{dt})\mathbf{\tilde{v}}$ and $Q_{k}(\tfrac{d}{dt})\mathbf{\hat{i}} = P_{k}(\tfrac{d}{dt})\mathbf{\hat{v}}$. 
}
\end{rem}

\bibliographystyle{ifacconf-harvard}        
\bibliography{recip}

\end{document}